\documentclass[twocolumn,floatfix,aps,prb]{revtex4}
\usepackage{epsfig,subfigure,amssymb,amscd,amsmath,graphicx}

\def\be{\begin{equation}}
\def\ee{\end{equation}}
\def\ba{\begin{eqnarray}}
\def\ea{\end{eqnarray}}

\def\nl{\nonumber \\}

\def\d{\delta}

\newcommand{\braket}[2]{\left\langle #1 | #2 \right\rangle}

\begin{document}

\title{Quasiparticles and excitons for the Pfaffian quantum Hall state}

\author{Ivan~D.~Rodriguez$^{1}$, A.~Sterdyniak$^{2}$\footnote{IDR and AS should be considered joint first authors of this paper}, M.~Hermanns$^{3}$, J.~K.~Slingerland$^{1,4}$, N.~Regnault$^{2}$}

\affiliation{
$^{1}$ Department   of  Mathematical  Physics,  National University of Ireland, Maynooth, Ireland.\\
$^{2}$ Laboratoire Pierre Aigrain, ENS and CNRS, 24 rue Lhomond, 75005 Paris, France.  \\  
$^{3}$ Department of Physics, Princeton University, Princeton, NJ 08544\\
$^{4}$ Dublin Institute for Advanced  Studies, School of Theoretical  Physics, 10 Burlington Rd, Dublin, Ireland.}

\begin{abstract}
\noindent
We propose trial wave functions for quasiparticle and exciton excitations of the Moore-Read Pfaffian fractional quantum Hall states, both for bosons and for fermions, and study these numerically. Our construction of trial wave functions employs a picture of the bosonic Moore-Read state as a symmetrized double layer composite fermion state. We obtain the number of independent angular momentum multiplets of quasiparticle and exciton trial states for systems of up to $20$ electrons. We find that the counting for quasielectrons at large angular momentum on the sphere matches that expected from the CFT which describes the Moore-Read state's boundary theory. In particular, the counting for quasielectrons is the same as for quasiholes, in accordance with the idea that the CFT describing both sides of the FQH plateau should be the same. We also show that our trial wave functions have good overlaps with exact wave functions obtained using various interactions, including second Landau level Coulomb interactions and the $3$-body delta interaction for which the Pfaffian states and their quasiholes are exact ground states. We discuss how these results relate to recent work by Sreejith et al.~on a similar set of trial wave functions for excitations over the Pfaffian state as well as to earlier work by Hansson et al., which has produced trial wave functions for quasiparticles based on conformal field theory methods and by Bernevig and Haldane, which produced trial wave functions based on clustering properties and `squeezing'. 
\end{abstract}

\maketitle

\section {INTRODUCTION}
\label{sec:Introduction}
The fractional quantum Hall plateau observed\cite{Willett87,Pan99} at filling $\nu=\frac{5}{2}$ has recently been at the center of much excitement, because it is expected that the elementary charged excitations of this state may be non-Abelian anyons.\cite{Moore91,Greiter92,Nayak96c,Read96,Bonderson11_PRB_83} Moreover, manipulation of these anyons potentially represents an avenue to topologically fault tolerant and hence scalable quantum computation\cite{Kitaev97,Freedman98,Preskill98,Nayak08}. Various experiments have already observed important signatures of these excitations, such as their fractionalized charge\cite{Dolev08_Nature} and tantalizing interferometric properties\cite{Willett09,Willett10}, which may provide a smoking gun detection of non-Abelian anyonic statistics\cite{DasSarma05,Bonderson06a,Stern06a,Bonderson06b}.  

Moore and Read's Pfaffian wave function~\cite{Moore91}, and its particle-hole conjugate, the anti-Pfaffian\cite{Lee07,Levin07} remain the leading candidates for the description of the electronic ground state at $\nu=\frac{5}{2}$. However, since Hall plateaus are probed through their excitations, it is of great importance to understand not only the ground state, but also the spectrum of low lying excitations which are naturally associated with Moore and Read's proposal. Much effort has been devoted to understanding the quasihole excitations, that is, the low energy states which are appropriate in situations where there is more magnetic flux piercing the sample than at the center of the Hall plateau (or equivalently, the electron density is lowered compared to the center of the plateau). A natural set of candidate wave functions exists for these, namely the exact zero energy ground state wave functions of the model $3$-body Hamiltonian introduced in Refs.~\onlinecite{PhysRevLett.66.3205,Greiter92}. This Hamiltonian has the property that the Pfaffian state is its highest density (or lowest angular momentum) zero energy state. At lower densities (when quasiholes are present), it has multiple zero energy states and one may conjecture that these states are good trial wave functions for the low energy Coulomb spectrum. In fact, the trial wave functions for \emph{localized} quasiholes for the MR-states, which are constructed using correlators in conformal field theory (CFT), are automatically zero energy eigenstates of the $3$-body Hamiltonian\cite{Read96}, and hence coherent superpositions of its zero modes, because this Hamiltonian encodes properties of the operator product expansion in the CFT. Numerical studies\cite{Read96,Toke2007504,Regnault07} on small systems have indeed found that the zero modes of the $3$-body interaction provide a reasonable description of the low energy spectrum.

Much progress has also been made in gaining a analytical understanding of the zero energy states of this model Hamiltonian and its generalizations with $k$-body interactions, which play a similar role for the Read-Rezayi series of states~\cite{Read99}. Notably, it is known exactly how many independent zero energy states exist at any number of particles $N$ and flux quanta $N_{\phi}$ and even how many of these states exist with any given angular momentum~\cite{Ardonne02a,Read06}.  These countings are an important fingerprint of statistical properties of the excitations and of the conformal field theory (CFT) which describes the edge of a system with boundary in the thermodynamic limit. 

Not nearly as much is known about the quasielectron excitations, which occur at higher density or lower magnetic field, and about the neutral excitations (excitons), which can be viewed as combinations of quasiholes and quasielectrons. Clearly one may still conjecture (as done implicitly e.g.~in Ref.~\onlinecite{Read96}) that the low energy states of the $3$-body Hamiltonian are good trial wave functions, but these are now no longer exact zero modes, and no exact expression is known for them. The aim of this paper is to propose and study an alternative set of trial wave functions for the neutral and charged excitations of the fermionic and bosonic Moore-Read Pfaffian states, for which we can write down an explicit analytical form. The fermionic wave functions we propose are very similar in construction to those recently studied by Sreejith, T{\H o}ke,W{\'o}js and Jain in Refs.~\onlinecite{Sreejith11,Sreejith11_PRL_107}, which appeared while the present work was being written. In fact, though there are subtle differences between our construction and theirs (see Section~\ref{sec:traditional_method}), which can cause differences in the quality of the approximation of real systems, we believe the two sets of wave functions should be able to describe the same fractional quantum Hall universality class. Nevertheless, our results on the counting of multiplets of quasielectrons are markedly different from those presented in Ref.~\onlinecite{Sreejith11_PRL_107}. There it was stated that the counting for quasielectrons is different from that for quasiholes and the possibility was raised that the quasielectrons might have different braiding properties from quasiholes. However, we find that quasielectrons satisfy the same universal counting properties as quasiholes, suggesting that they also enjoy the same braiding properties. 

Our construction of trial wave functions is based on the idea that we can view the bosonic Moore-Read state as a double layer\cite{Cappelli01} of bosonic Laughlin 1/2 states--- the fermionic MR ground state can be obtained by multiplying with a Jastrow factor involving all the particles.  A construction of localized quasiholes from the double layer representation of the MR wave function was already given in Ref.~\onlinecite{Cappelli01}. Here, we generate trial wave functions for all excitations of the bosonic and fermionic MR-states by constructing, in each layer, all the possible excitations (excitons, quasiholes, and quasielectrons) over the $\nu=1/2$ Laughlin ground states. The latter are well-understood in terms of the composite fermion (CF) theory \cite{Jain89}, that is, by creating quasielectrons and/or quasihole excitations in the integer quantum Hall effect of the CF at effective filling fraction $\nu^{*}=1$. We calculate the numbers of independent angular momentum multiplets of such states which exist at numerically accessible $N$ and $N_{\phi}$, providing evidence that our construction yields quasielectrons with the same type of non-Abelian statistics and edge CFT as the quasiholes. We also compare the trial wave functions to the low energy eigenfunctions of the 3-body Hamiltonian and the second Landau level Coulomb Hamiltonian, showing directly that they indeed are good candidates to describe the excitations of the MR Pfaffian state. 

Other candidates for quasielectron and/or exciton trial wave functions, not directly based on composite bosons or fermions, have been proposed previously by Hansson, Hermanns, Regnault and Viefers\cite{Hansson09,Hansson09a} and by Bernevig and Haldane\cite{Bernevig06}. The construction in Refs.~\onlinecite{Hansson09,Hansson09a} is based on CFT and looks superficially very different to ours, but nevertheless, we expect that the quasiparticle wave functions presented there are the maximally  localized (coherent) states, which can be produced from the states we propose here. The construction in Ref.~\onlinecite{Bernevig06} determines the trial wave functions for quasielectrons by requiring that they vanish when certain patterns of clusters of electrons are formed and also that they be dominated by certain root configurations. We review both constructions in some detail in the rest of the paper and comment on the  similarities and differences to our own construction. 

We have focused our numerical tests on systems with an even number of electrons, in part because these systems exhibit a unique incompressible ground state, making quasielectrons and excitons clearly defined. Recently a number of works has also appeared which study the band of low energy states which appears in systems at $\nu=\frac{5}{2}$ when the number of electrons is odd\cite{Bonderson11,Moller11,Sreejith11}. In particular, the paper by Sreejith et al.\cite{Sreejith11}  employs trial wave functions for the states in this band which are based on a double layer CF system. Further recent work which focuses on the properties of excitons at $\nu=\frac{5}{2}$, and particularly on the roton minimum, includes Refs.~\onlinecite{Simion10,Wright11}.

\paragraph*{Outline of the paper.}
In Section~\ref{sec:MR-review}, we give an overview of the Moore-Read state and of the quasielectron constructions of Hansson et al.~and Bernevig and Haldane. 
In Section~\ref{sec:CF-review}, we give a quick review of the construction of excitations over composite fermion or composite boson ground states. 
In Section~\ref{sec:traditional_method}, we describe our own trial wave functions for quasielectrons and excitons and explain how they can be numerically evaluated and studied both by real space Monte Carlo methods and by Fock space methods (using eigenstates of angular momentum), which allow us to work at machine precision.  In particular, we give details on the calculation of overlaps and of the number of independent trial wave functions for given $N$, $N_{\phi}$ and number of excitons or quasielectrons. In this section, we also give a detailed explanation of the relation between our trial wave functions and those of Ref.~\onlinecite{Sreejith11_PRL_107}.
In Section~\ref{sec:numerical_results}, we present our numerical results, which include state counting of quasielectrons and excitons and overlaps between our sets of wave functions for bosons and fermions and the low lying states in the exact spectra obtained for bosons and fermions with three-body hardcore interactions, two-body hardcore interactions and second Landau level Coulomb interactions (with a slight shift of the pseudopotential $V_1$ to obtain a stable Pfaffian state). 
We note that our results for quasielectron counting are consistent with the idea that the conformal field theory on the edge of a disk containing MR-type FQH-liquid should be the same on both sides of the plateau. 
Finally in Section~\ref{sec:discussion}, we critically examine our results and discuss potential future developments. 



\section{Quasielectron constructions over the Moore-Read ground state}
\label{sec:MR-review}

Here, we briefly review the MR state and its quasihole excitations and discuss two existing constructions of trial wave functions for quasielectrons and excitons --- one using the language of CFT, conjectured by Hansson, Hermanns, Regnault and Viefers \cite{Hansson09,Hansson09a}, and one conjectured by Bernevig and Haldane, who define their trial wave functions by their vanishing properties \cite{Bernevig06}.  Although the language used to describe the model states is very different in the different approaches, the ground state and quasihole state wave functions are identical. However, each approach has a 'natural' extension towards quasielectrons, leading to distinct, but related model wave functions. We will comment on their relation and differences in Sections \ref{sec:traditional_method} and \ref{sec:numerical_results}.

In this section, we will focus on the simplest case: the bosonic MR state at filling $\nu=1$.  In an abuse of language, we still use the words electron and 'quasielectron', even though the system is made up of bosons. 


\subsection{Ground state and quasihole excitations}
\label{sec:CFTreview}
Let us start by reviewing some important properties of the MR ground state and its quasihole excitations. It was noted early on in Ref.~\onlinecite{Fubini91} that Laughlin model states as well  as their quasihole excitations can be written as correlation functions, where the particles are represented by CFT operators. Moore and Read generalized this approach\cite{Moore91}, and proposed a model wave function  based on the Ising CFT for the fermionic FQHE state at filling $2+1/2$. In the following, we focus on the bosonic version of this state--- the MR Pfaffian state at filling  $\nu=1$:

\begin{align}
\label{CFT-M-R-GS}
\Psi_{\rm Pf}(z_1,\ldots,z_N)={\rm Pf}\left( \frac{1}{z_i-z_j} \right)\cdot \prod_{i<j}(z_i-z_j). 
\end{align}
Pf($A$) denotes the Pfaffian of the skew-symmetric matrix $A$. It is defined by 
\begin{align}
{\rm Pf}\left(A\right)=&\sum_{\sigma} \epsilon_{\sigma}
A_{\sigma(1)\sigma(2)}
A_{\sigma(3)\sigma(4)}...A_{\sigma(N-1)\sigma(N)}, 
\label{pfaffiandefinition}
\end{align}
where the sum runs over all permutations $\sigma$ of the $N$ indices, and  $\epsilon_{\sigma}$ is the signature of the permutation.

The model state \eqref{CFT-M-R-GS} is the densest (lowest degree) zero-energy state of  $\mathcal{H}_{B}^{(3)}$--- the hardcore three-body Hamiltonian ($k=2$),
\ba
\mathcal{H}_{B}^{(k+1)} = \sum_{i_1<\dots<i_{k+1}} \prod_{j=1}^k \d^2 (z_{i_{j}}-z_{i_{j+1}}).
\label{3bHam}
\ea
In particular, the bosonic MR Pfaffian state vanishes as the second power of the difference between coordinates when three particles come to the  same position. More precisely,
\ba
\Psi_{\rm{Pf}}(z_1=z_2,z_3,..,z_N) \sim \prod_{i = 3}^N (z_1 - z_i)^2\, .
\label{pfaffian_vp}
\ea
This vanishing property is a particular case of the more general $(k,r)$ vanishing properties, where the polynomials vanish as the $r$th power when $k+1$ particles come to the same point: \cite{Bernevig08_prl_101}
\ba
\Psi_N^{(k,r)}(\underbrace{z_1 = z_2 =\ldots=z_k}_z,z_{k+1},\ldots,z_N) = \nl \prod_{i = k+1}^N (z - z_i)^r\Psi_{N-k}^{(k,r)}(z_{k+1},\ldots,z_N)\, .
\label{general_vp}
\ea

Note that for fermionic systems there are ultralocal Hamiltonians $\mathcal{H}^{(k+1)}_{F}$, similar to the $\mathcal{H}^{(k+1)}_{B}$, which implement vanishing properties for the wave functions so that, after division of the wave function by a Jastrow factor (which is always possible for a fermionic wave function), the resulting function still vanishes when $k+1$ of the coordinates are equal. 

The quasihole state manifold of the MR state is spanned by less dense (higher degree) polynomials that satisfy the vanishing conditions \eqref{pfaffian_vp}. These quasihole states are in fact ground states (i.e.~zero-modes) of the three-body Hamiltonian \eqref{3bHam}, albeit at a higher number of flux quanta $N_\phi$ than the Pfaffian state (\ref{CFT-M-R-GS}). Nevertheless, we will call them excitations, because, when perturbing away from the model Hamiltonian towards more realistic Hamiltonians, the degeneracy between these states is split and the resulting band should give a good description of the low-energy sector of the more realistic system.

One may choose a basis for the space of quasihole states which consists of eigenstates of the total angular momentum operator $\hat L$ and the angular momentum along the $z$-direction, $\hat L_z$.  The quasihole counting, i.e.~the number of basis states (or equivalently the number of multiplets) at each angular momentum, is a fingerprint of the topological order of the model state. For the MR Pfaffian states, it was explained in Ref.~\onlinecite{Read96} how to calculate the number of quasiholes states $\mathcal{N}(N,n_{qh},l_z)$ for $N$ particles and $n_{qh}$ quasiholes with $\hat L_z$ eigenvalue $l_z$. Formulas for $\mathcal{N}(N,n_{qh},l_z)$ for the entire family of Read-Rezayi states were obtained using CFT methods in Ref.~\onlinecite{Ardonne02a} (see also Ref.~\onlinecite{Ardonne05}) and by direct  counting of polynomials with the required vanishing properties and degree restrictions in Ref.~\onlinecite{Read06}. The number of multiplets at $L\geq 0$ can always be easily found from the numbers of states at given $\hat L_z$ eigenvalue to be $\mathcal{N}(N,n_{qh},L)-\mathcal{N}(N,n_{qh},L+1)$. 
To give detailed results on $\mathcal{N}(N,n_{qh},l_z)$, let us first define the q-binomial $\genfrac{[}{]}{0pt}{}{a}{b}$ by
\begin{align}
\label{qe:qbinom}
\genfrac{[}{]}{0pt}{}{a}{b}
=\left\{\begin{array}{cl}
\frac{(q)_a}{(q)_{a-b} (q)_b} & \mbox{ for } a,b\in \mathbb{N}, \, a\geq b \\
0& \mbox{ otherwise}
\end{array} \right.
\end{align}
where $(q)_m=\prod_{j=1}^m (1+q^j)$.
Following Ref. \onlinecite{Ardonne02a} the generating function of  $\mathcal{N}(N,n_{qh},l_z)$ can then be written as:
\begin{multline}
\label{qhole-count}
\sum_{l_z=-Nn/4 }^{Nn/4} \mathcal{N}(N,n,l_z) \cdot q^{l_z} =q^{-(2N+n_{qh})N/4}\\
\times\sum_{a=0}^{N/2} q^{N^2-2aN+2a^2}
\genfrac{[}{]}{0pt}{}{n_{qh}/2}{N-2a}
\genfrac{[}{]}{0pt}{}{n_{qh}+a}{a}
\end{multline}
For the first $\min[N/2,n_{qh}]$ angular momentum multiplets (counted from the highest) the quasihole counting is identical to the edge mode counting that is expected in the thermodynamic limit in the disc geometry. This is reasonable: the sphere and the disc are 
connected via a stereographic projection. The south pole of the sphere is mapped to the origin, while the North pole is mapped to the edge of the disc. Thus, we expect the state counting at high angular momenta on the sphere to correspond to the edge counting in the disc geometry. In Section \ref{sec:numerical_results}, we will see that the situation is similar for the quasielectron state counting; the high angular momentum counting is identical to the edge counting, even though finite size corrections to the counting appear earlier than for the quasihole case. 

\subsection{Quasielectron construction based on CFT}
\label{sec:CFTqe}
We now discuss the construction of trial wave functions for quasielectrons by Hansson et al.\cite{Hansson09} using CFT. Note that this construction focused on \emph{localized} excitations. However, by expanding such wave functions in a basis of eigenstates of angular momentum, one may always obtain a set of trial wave functions for low energy excitations on a sphere, which can be compared to the set of trial wave functions to be presented in this paper.
The Pfaffian ground state, Eq.~\eqref{CFT-M-R-GS}, has a natural interpretation as a symmetrized wave function of two independent layers of Laughlin $\nu=1/2$ states:
\begin{multline}
\label{Pfaff-331}
\Psi_{\rm Pf}(z_1,\ldots,z_N) =\\
\mathcal{S}\left(\prod_{i<j=2}^{N/2} (z_i-z_j)^2 (z_{\frac{N}{2}+i}-z_{\frac{N}{2}+j})^2 \right).
\end{multline} 
Following Cappelli et al.\cite{Cappelli331,Cappelli01}, we may now observe that this state can be written as a CFT correlator using two independent bosonic fields: $\phi_c$, which is related to the filling fraction and thus to the electric charge, and $\phi_l$, which distinguishes the two different layers. 
In the CFT we associate an operator $V$ with the electron, given by
\begin{align}
V(z)&=\cos[\phi_l(z)]e^{i\phi_c(z)} \nonumber\\
&= \frac{1}{2} (V_+(z)+V_-(z))\, ,
\end{align}
where $V_\pm(z)=\exp[\pm i \phi_l(z)+i\phi_c(z)]$ can be interpreted as electron operators in layer 1 ($+$) and layer 2 ($-$). 
The Pfaffian ground state can then be written as
\begin{align}
\Psi_{\rm Pf}(z_1,\ldots,z_N) &=
\langle \prod_{j=1}^N V(z_j) \mathcal{O}_{bg}\rangle \nonumber \\
&= \mathcal{S} \left[ \langle \prod_{j=1}^{N/2} V_+(z_j) \prod_{j=N/2+1}^N V_-(z_j)\,\mathcal{O}_{bg}\rangle\right]\, .
\end{align}
The homogeneous compensating  background charge operator $\mathcal{O}_{bg}$, see Ref.~\onlinecite{Moore91} for details,  is needed to render the correlation function charge neutral, and thus, non-zero. 
Note that it only contains the field $\phi_c$, but not the layer field $\phi_l$.  This means that non-zero correlation functions necessarily need to be charge-neutral in $\phi_l$. 
The background charge reproduces the correct Gaussian factor needed for a valid LLL wave function, but is otherwise of no importance for the remainder of this section.

Using this two-layer description, the non-abelian quasihole at position $\eta$ has to be described by two operators, 
\begin{align}
H_\pm (\eta)=\exp\left[\pm \frac{i}{2} \phi_l(\eta)+\frac{i}{2} \phi_c(\eta)\right].
\end{align}
 Without the symmetrization procedure, $H_+$ and $H_-$ would be Abelian (Laughlin-type) quasiholes in each layer. Because of charge-neutrality, the quasiholes can only be inserted in pairs. 
 The non-abelian nature of the quasiholes manifests itself in a topological degeneracy of $2^{n-1}$ for $2n$ localized quasiholes. 
 In Moore and Read's original description using the Ising CFT, this degeneracy originates in the two possible fusion channels of the CFT operator describing the non-Abelian quasihole. 
 In the two-layer description, the degeneracy comes from the possible 'distribution' of the $2n$ quasihole positions $\eta_1, \ldots, \eta_{2n}$ over the two layers. 
 This naively over-counts the number of quasihole states. 
 However, using techniques described in Ref.~\onlinecite{Nayak96c}, one can show that not all of them are linearly independent and that suitable linear combinations reproduce the localized quasihole wave functions obtained from the Ising description. 
 For instance, for four quasiholes, there are 2 linearly independent states:
\begin{align}
\Psi_{MR}^{4qh,1}&= \langle H_+(\eta_1)H_+(\eta_2)H_-(\eta_3)H_-(\eta_4)\prod_{j=1}^{N} V(z_j) \mathcal{O}_{bg}\rangle \nonumber\\
\Psi_{MR }^{4qh,2}&= \langle H_+(\eta_1)H_-(\eta_2)H_+(\eta_3)H_-(\eta_4)\prod_{j=1}^{N} V(z_j) \mathcal{O}_{bg}\rangle\, . \nonumber
\end{align}
Expanding the localized quasihole states in angular momentum eigenstates reproduces the zero-energy multiplets found by diagonalizing the three-body Hamiltonian \eqref{3bHam}.

In Ref. \onlinecite{Hansson09}, Hansson et al.~introduce quasielectron operators which play a similar role in the definition of trial wave functions with quasielectrons as the operators $H_{\pm}$ do for  wave functions with quasiholes. The guiding principle used there to construct these quasielectron operators is to view the quasielectron as the antiparticle of the quasihole. 
However, the operators $H^{-1}_{\pm}(\eta)$ are obviously not good candidate quasielectron operators, because they produce singularities in the electron coordinates. Instead, the authors of Ref.~\onlinecite{Hansson09} constructed well-defined, regularized operators $\mathcal{P}_{\pm}$ with the same long range properties as $H^{-1}_{\pm}(\eta)$; For more details on the regularization, see Ref.~\onlinecite{Hansson09a}. 
The operators, $\mathcal{P}_{\pm}$ can be interpreted as abelian quasielectrons in the $\pm$ layers. As was the case for quasiholes, explicit symmetrization is essential for the non-abelian properties. It was shown in Ref.~\onlinecite{Hansson09a} that localized quasielectrons have the same topological multiplicity as localized quasiholes, that is, for $2n$ localized quasielectrons, there are $2^{n-1}$ linearly independent candidate wave functions. For instance, the 2-quasielectron candidate wave function is unique and given by (see Ref.~\onlinecite{Hansson09a} for the 4-quasielectron candidate):
\begin{multline}
\label{eq:2qeCFT}
\Psi_{MR}^{2qe}(\{z_j\}) =\langle \mathcal{P}_+(\eta_1)\mathcal{P}_-(\eta_2) \prod_{j=1}^N V(z_j)\rangle\\
=\mathcal{S}\left[ e^{(\bar\eta_1 z_1+\bar\eta_2 z_{N/2+1})/8\ell^2} \right.
\left(\partial_{1} \prod_{j=2}^{N/2} (z_1-z_j)\right) \prod_{2\leq i<j}^{N/2} (z_i-z_j)^2
\\ 
\times \left(\partial_{\frac{N}{2}+1}\prod_{j=\frac{N}{2}+2}^{N} (z_{\frac{N}{2}+1}-z_j)\right)
 \left. \prod_{N/2+2\leq i<j}^{N} (z_i-z_j)^2\right]\, ,
\end{multline}
where we abbreviated $\partial_j\equiv \partial_{z_j}$. This method can be applied for an arbitrary number of quasielectrons and/or quasiholes by inserting the appropriate operators in the first line of \eqref{eq:2qeCFT}.


\subsection{Quasielectron construction based on vanishing conditions}
\label{sec:Jacks}
\label{sec:BH}

Bernevig and Haldane conjectured a quasielectron construction by imposing vanishing and clustering properties on the candidate wave functions\cite{Bernevig06}.  Before going into the details of their construction, let us review some important background material. 
The bosonic RR model states, as well as their quasihole excitations, are uniquely defined by their vanishing properties. 
In the case of the bosonic MR state, the ground state is the lowest degree symmetric polynomial that vanishes when 3 particles are at the same positions. 
Higher degree polynomials, obeying this vanishing property, span the quasihole state manifold. 

The single-particle states in the LLL, 
\[ \phi_n(z)=(2 \pi n! 2^n)^{-\frac{1}{2}} z^n e^{-|z|^2/4},\]
are  eigenstates of the angular momentum operator, $\hat L_z$ with eigenvalues which are just proportional to powers of the complex coordinate $z$. Hence, there is a basis for the many-body states which consists of symmetrized monomials $m_\lambda$, where $\lambda$ is a partition of the total angular momentum $n_{t}$.
Alternatively, one can label a monomial by its corresponding occupation number configuration $n_\lambda=\{n_j(\lambda),\, j=0,1,\ldots, N_\phi\}$, where $n_j(\lambda)$ is the occupation number of the single-particle state with angular momentum $j$. 
A set of partitions may always be partially ordered by dominance, denoted by '$>$'. A partition $\mu$ dominates another partition $\nu$ ($\mu>\nu$) if the latter can be obtained by successive squeezing operations on $\mu$. Squeezing is a two-particle operation that changes the angular momenta of two particles from $j_1$ and $j_2$ to $j_1'$ and $j_2'$, such that $j_1<j_1'\leq j_2'<j_2$ and total angular momentum is conserved. 

It has been realized recently~\cite{Bernevig08_prl_100,Bernevig08_prl_101}, that many fractional quantum Hall trial wave functions, in particular the ground states and quasiholes of the Read Rezayi series, can be written as Jack polynomials. This means in particular that they have non-zero coefficients only for a small subset of the monomials which span the full Hilbert space. In fact, for each of these wave functions, there is a special partition $\lambda_0$, called the \emph{root partition}, such that $m_{\lambda_0}$ has non-zero weight and $\lambda_0$ dominates any other partition present in the expansion:
\begin{align}
\label{eq:monExp}
P_{\lambda_0} &=m_{\lambda_0}+\sum_{\mu<\lambda_0} v_{\lambda,\mu} m_\mu\, .
\end{align}
Jack polynomials satisfy further properties in addition to the fact that they have a nontrivial root partition. In particular, the coefficients of the nonzero monomials in the expansion of a Jack may be obtained from a recursion relation.\cite{bernevig-09prl206801,stanley89advm76} However, in describing Bernevig and Haldane's quasielectron construction, we will use wave functions which have a given root partition and additional vanishing properties, but which are not necessarily Jacks. 
 
Because the MR ground state is the lowest degree polynomial that satisfies the vanishing properties ~(\ref{pfaffian_vp}), and inserting quasielectrons necessarily involves lowering the total degree, Bernevig and Haldane suggested that the quasielectron polynomials are defined by modified vanishing conditions. They conjectured root configurations and vanishing  conditions for two types of quasielectrons which they call Abelian quasielectrons and non-Abelian quasielectrons. Abelian quasielectrons carry a full flux quantum (that is, in creating Abelian quasielectrons, one must lower the electric flux by one quantum per quasielectron), while non-Abelian quasielectrons carry only half of a flux quantum, like the non-Abelian anyonic half-flux quasiholes of the Pfaffian state.  
In their paper, Ref.~\onlinecite{Bernevig06}, Bernevig and Haldane focus on systems with any number of Abelian quasielectrons (all localized near the same position) and systems with a single non-Abelian quasihole and a single non-Abelian quasielectron (these are really excitons). 
Let us start by reviewing the vanishing  and clustering conditions for the Abelian quasielectrons of the MR state, because they are slightly simpler. 
In the following, the expression 'forming a cluster of $n$ particles' denotes that $n$ particles are at the same positions. 
The requirements on an $s-$Abelian quasielectron state (which is a state with $s$ Abelian quasielectrons localized near the same position) are that it vanishes when $s+1$ clusters of 4 particles are formed, and it vanishes  when one cluster of $2s+3$ particles is formed as the $(s+2)$th power of the difference between coordinates. In the special case of $s=1$ this becomes 
\begin{align}
\label{eq:clusteringA}
&P(z_1,z_1,z_1,z_1,z_2,z_2,z_2,z_2, z_9,\ldots, z_N)=0\nonumber\\
&P(z_1,\ldots, z_1, z_6,\ldots,z_N)\sim \prod_{j=6}^N (z_1-z_j)^3
\end{align}
with the root partition $\{4\,0\,0\,2\,0\,2\ldots 02\}$.

The root configurations and vanishing conditions for a non-Abelian quasielectron-quasihole pair for the RR $Z_k$ states are also given explcitly in Ref.~\onlinecite{Bernevig06}. 
The quasielectron-quasihole candidate states for the MR state form angular momentum multiplets  $L=2,3,\ldots,N/2$. The highest weight states of these multiplets are defined by the fact that they have the following root configurations
\begin{align}
\label{eq:rootNA}
L=N/2;&\hspace{0.4cm} \{3\, 0\, 1\, 1\, 1\, 1\ldots \, 1\, 1\, 1\}\nonumber\\
L=N/2-1; &\hspace{0.4cm} \{3\, 0\, 1\, 1\, 1\, 1\ldots \, 1\, 0\, 2\}\nonumber\\
\vdots &\nonumber\\
L=2;&\hspace{0.4cm} \{3\, 0\, 1\, 0 \, 2 \, 0 \ldots 2\, 0\, 2\}
\end{align}
and by requiring in addition that they vanish for 2 clusters of 3 particles and for a single cluster of 4 particles:
\begin{align}
\label{eq:clusteringNA}
P(z_1, z_1, z_1, z_2, z_2, z_2, z_7,\ldots, z_N)=0\nonumber\\
P(z_1, \ldots, z_1, z_5, \ldots, z_N)=0.
\end{align}
These conditions can be generalized to several quasihole-quasielectron pairs and also to an even number of non-Abelian quasielectrons\cite{bernevigPC}. 
The quasielectron-quasihole states satisfy a further vanishing property, namely
\begin{equation}
\begin{array}{l}
\label{eq:extravanish}
P(z_1,z_1,z_1,z_2,z_2,z_6,\ldots,z_{N})\\
~~~~~~~~~~~\sim (z_1-z_2)^{3}\prod_{i=6}^{N}(z_1-z_i)^2(z_2-z_i)^2
\end{array}
\end{equation}
This condition, in combination with the second condition in ~(\ref{eq:clusteringNA}) also uniquely determines the space of quasihole-quasielectron pair wave functions.
We compare our trial wave functions for excitons and quasielectrons to those proposed by Bernevig and Haldane in Section~\ref{sec:BHcomp}.


\section {JAIN COMPOSITE FERMION PICTURE}
\label{sec:CF-review}

\subsection{CF quasielectrons and excitons}

In this section we give a short review of the treatment of excitations over fractional quantum Hall plateaus based on composite fermions (CF), as introduced by Jain~\cite{Jain89}. Composite fermions provide an attractive physical picture and very successful trial wave functions for the most prominent filling fractions in the LLL. For a much more in depth review and extensive references, see for instance Ref.~\onlinecite{Jain_CF}. 

Jain conjectured that states of strongly interacting electrons can be understood in terms of states of non-interacting or weakly-interacting  composite particles, called composite fermions. 
A composite fermion consists of a fermion (boson) and an even (odd) number $m$ of vortices. 
When moving in a magnetic field, the attached vortices generate a Berry phase that partly cancels the Aharonov-Bohm phase. 
Thus, the CFs behave as if they were subject to a reduced magnetic field, ${\bf B}^* ={\bf B} - m \phi_0\rho$ with $\rho$ being the density and $\phi_0$ the magnetic flux quantum. 
The reduced magnetic field ${\bf B}^*$ gives rise to Landau like levels,  called $\Lambda$ levels ($\Lambda$Ls) in the following, which are separated by an effective CF cyclotron energy $\hbar w_c^*$. 

\begin{figure}[ht!]
\begin{center}
\includegraphics[width= 8cm]{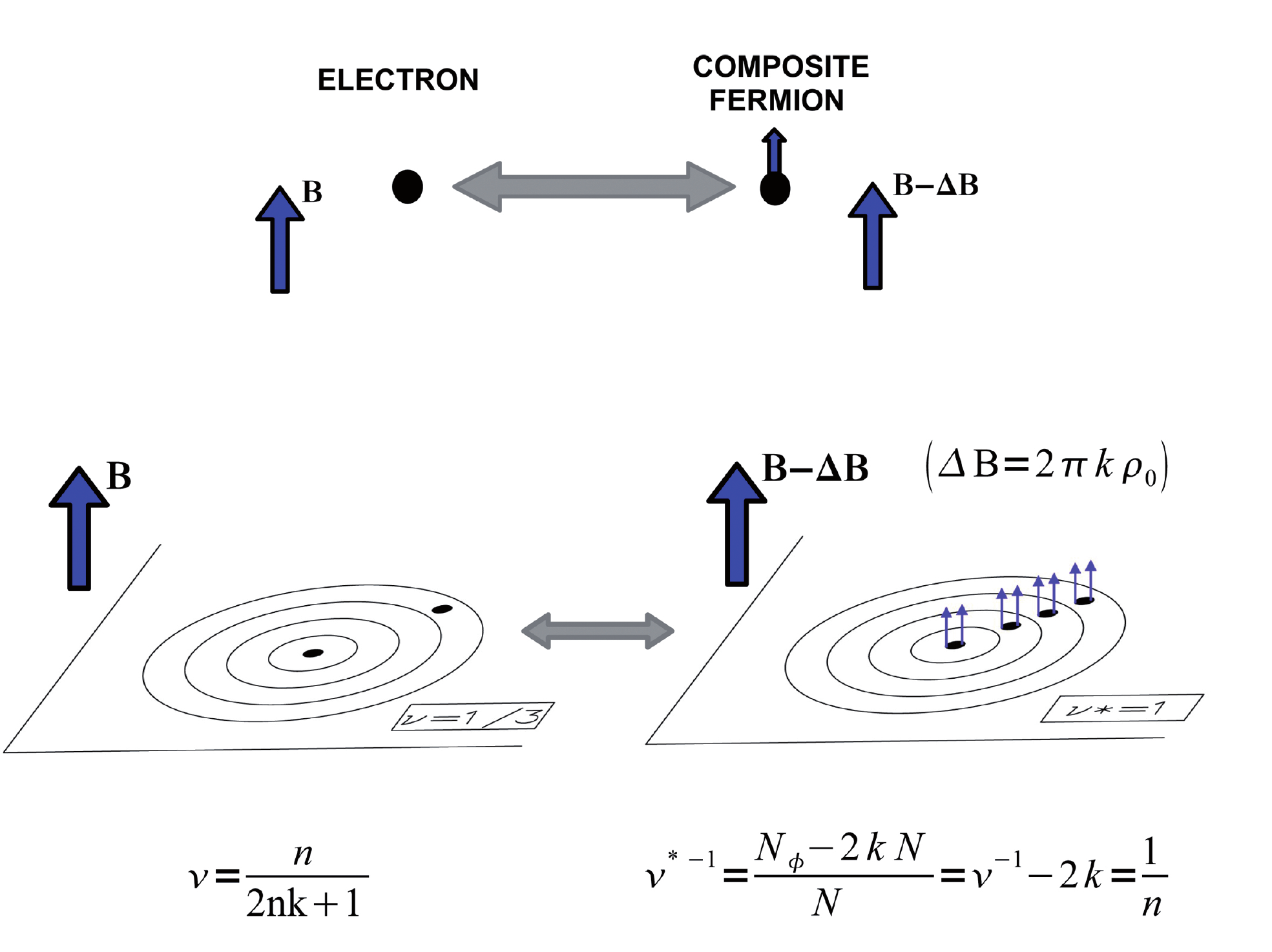}
\caption{Sketch of the composite fermion mapping in the case of two vortices attached ($m = 2$ in Eq. \eqref{Jain}).}
\label{fig1}
\end{center}
\end{figure}

Based on this interpretation, Jain proposed a generalization of the Laughlin wave function, describing the FQHE at filling fraction $\nu=\frac{n}{nm+1}$ effectively as an integer quantum Hall (IQH) state of CFs at filling $\nu^{*}=n$:
\begin{align}
\Psi^{CF}_{\frac{n}{nm+1}}(z_1,\ldots,z_N)=\mathcal{P}_{LLL}\left[\prod_{i<j}^N (z_i-z_j)^{m} \phi_{n} (z_1,...,z_N)\right].
\label{Jain}
\end{align}
Here, $\phi_{n} (z_1,...,z_N)$ is the IQH ground state wave function with $\nu^* = n$ completely filled $\Lambda$Ls. The $m$ Jastrow factors $\prod_{i<j}^N (z_i-z_j)^{m}$ attach $m$ vortices to each particle,  and $\mathcal{P}_{LLL}$ projects the wave function on the lowest Landau level (LLL). 
In the particular case $n=1$, the Jain state (\ref{Jain}) is identical to the Laughlin wave function for filling $\nu=\frac{1}{m+1}$. 

An important property of the CF picture is that it not only describes the ground states accurately, but also gives a very good description of the low-energy excitations, both neutral (excitons) and charged (quasiholes and quasielectrons). 
The trial wave functions for excitations are obtained by creating excitations in the $\nu^*=n$ IQH state of the CFs (again, see {\it e.g.} Ref. \onlinecite{Jain_CF} for a detailed review). Here, we don't present  the method in full generality, but rather give some representative examples.   

\begin{figure}[ht]
\begin{center}
\includegraphics[width= 8cm]{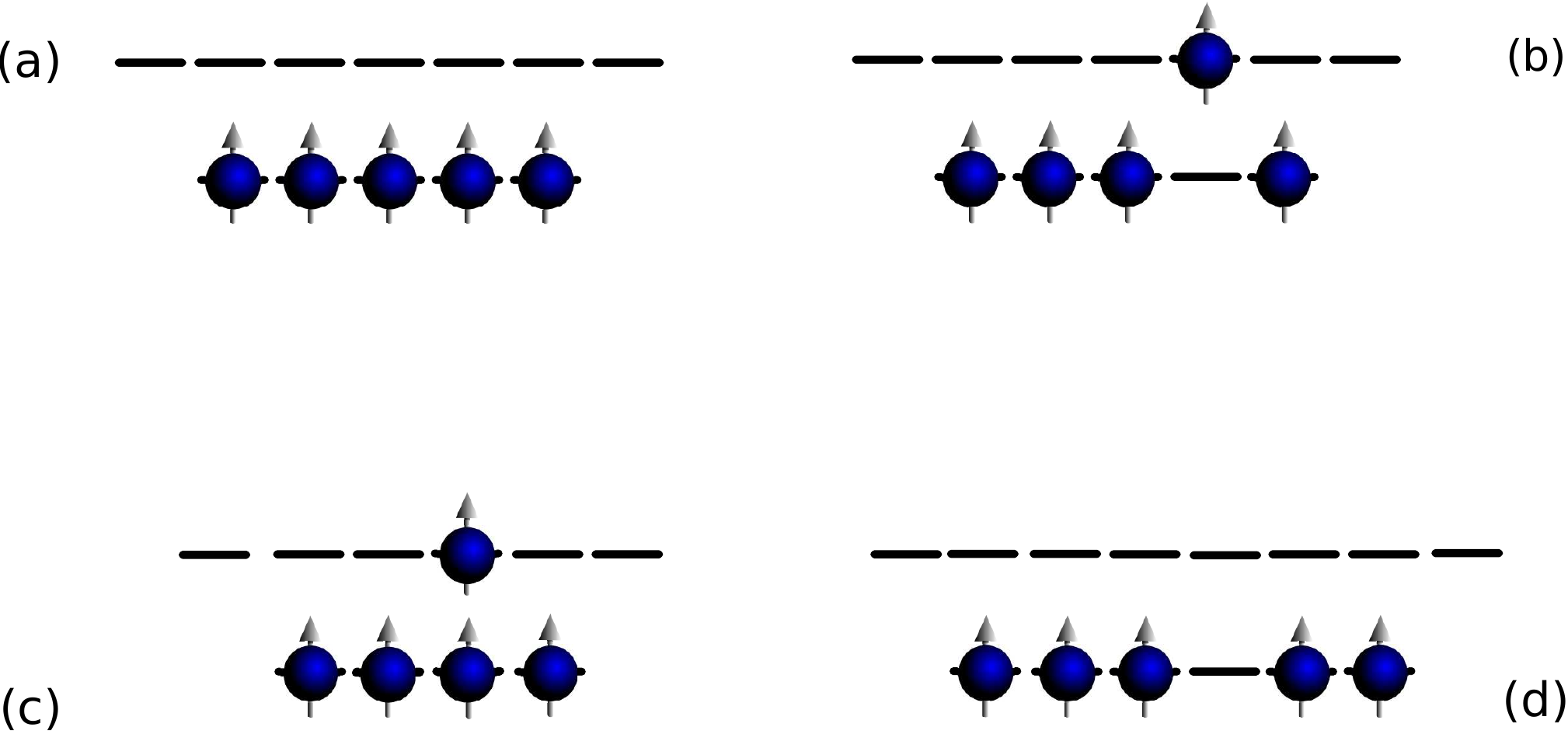}
\vspace*{5mm}
\caption{(a) The bosonic $\nu = 1/2$ Laughlin state in the composite fermion picture. (b) An exciton over this state, with one CF in the second $\Lambda$L and one hole in the L$\Lambda$L. (c) A quasielectron excitation of the Laughlin state: as the number of flux quanta is decreased by one unit, one CF has to occupy the second $\Lambda$L. (d) A quasihole excitation of the Laughlin state: as the number of flux  quanta is increased by one unit, there is one hole in the L$\Lambda$L.}
\label{fig2}
\end{center}
\end{figure}

The first example consists in creating an exciton with energy one (in the effective CF cyclotron energy $\hbar\omega_c^{\star}$ unit) over the Laughlin state with filling fraction $\nu=\frac{1}{m+1}$ (which corresponds to taking $n=1$ in (\ref{Jain})). To create this excitation, a CF in $\phi_{1} (z_1,...,z_N)$ (see Eq.~(\ref{Jain})) is removed from the angular momentum $l$ state in the lowest $\Lambda$ level (L$\Lambda$L) and placed in the angular momentum $j$ state in the second $\Lambda$ level, thus leaving a hole in the L$\Lambda$L (see Fig. \ref{fig2}b). This creates an exciton with total $\hat L_z$ eigenvalue $j-l$,
\begin{multline}
 \Psi^{exc}_{CF}(z_1,..,z_N) =\\
  \mathcal{P}_{LLL} \left( \tilde{\phi}_{j-l} (z_1,\ldots,z_N, \bar z_1,\ldots,\bar z_N) \prod_{i<j} (z_i-z_j)^{m} \right)  
\label{Jainexc}
\end{multline}
with, up to overall normalization, 
\begin{multline}
\tilde{\phi}_{j-l} (z_1,\ldots,z_N,\bar z_1,\ldots,\bar z_N) = \epsilon^{i_1,\ldots,i_N}  \\
\times z_{i_1}^0 z_{i_2}^1 \dots z_{i_l}^{l-1}(\bar{z}_{i_{l+1}}z_{i_{l+1}}^{j+1})z_{i_{l+2}}^{l+1}\ldots z_{i_N}^{N-1} .
\label{Jainexc1}
\end{multline}
where $\epsilon^{i_1,\ldots,i_N}  $ is an antisymmetric tensor, and there is an implicit summation over repeated indices. The LLL projection can be implemented by putting all $\bar{z}$'s on the left and performing the following replacement in (\ref{Jainexc})\cite{PhysRevB.29.5617}:
\ba
\bar{z} \rightarrow 2 \frac{\partial}{\partial z} \ .
\label{Proj}
\ea
Therefore, $\tilde{\phi}_{1} (z_1,...,z_N, 2 \frac{\partial}{\partial z_1} ,..,2 \frac{\partial}{\partial z_N} )$ becomes an operator acting on the product $\prod_{i<j}^N (z_i-z_j)^{m}$.
In a similar way, we can generate charged excitations with angular momentum $l$. A quasielectron state with kinetic energy one is obtained by decreasing the number of quantum fluxes in $\phi_{1} (z_1,...,z_N)$ by one and placing a CF in the angular momentum $l$ orbital in the second  $\Lambda$ level (see Fig.~\ref{fig2}c) and finally projecting the wave function to the LLL using (\ref{Proj}).
A quasihole state over the Laughlin state is obtained by increasing the number of quantum fluxes in $\phi_{1} (z_1,...,z_N)$ by one. This creates a hole in the L$\Lambda$L with a given angular momentum (see Fig.~\ref{fig2}d).

\subsection{Vanishing properties of composite fermion states}
\label{sec:CFvanishing}

The vanishing properties of the states obtained from the CF picture are directly given by the index of the highest $\Lambda$ level occupied \cite{2011arXiv1105.5907S}: the bosonic states and the bosonic counterpart of the fermionic states (fermionic states divided by a Vandermonde determinant) constructed from $n$ $\Lambda$L satisfy equation (\ref{general_vp}) with  $k = n $ and $r = 2$. 
For instance, the bosonic Laughlin state and its quasihole states (see Figures \ref{fig2}a and \ref{fig2}d) reside completely in  the L$\Lambda$L; thus, they vanish when two particles are brought to the same point. 
The quasielectron and exciton states (see Figures \ref{fig2}b and \ref{fig2}c) involve states in the second $\Lambda$ level; therefore, they vanish when three particles are brought to the same point.
Moreover, the CF quasielectron states over a Laughlin state also have vanishing properties when multiple clusters are formed simultaneously: the 1-quasielectron states vanish when two clusters of two particles are formed\cite{Bernevig09_PRL_102}, whereas the 2-quasielectron states vanish when three clusters of two particles are formed\cite{bernevigUP}.

The trial wave functions for quasielectrons based on CFT obey the same vanishing conditions for single and multiple clusters as above.\footnote{Note that the vanishing properties do not define the CF-based trial wave functions uniquely in contrast to the approach by Bernevig and Haldane, where candidate wave function are defined by their vanishing properties and root configurations.}
 In fact, we expect the two approaches to yield identical vector spaces of trial wave functions for quasielectrons. In Ref.~\onlinecite{Hansson07long} this was shown for a single localized quasielectron (up to boundary effects that are absent in the spherical geometry). It should hold for several quasielectrons as well, when using the exact projection to the LLL\cite{PhysRevB.29.5617} in the CF construction. 
For neutral excitations of energy $\geq 2$, the two constructions can in principle differ, as the CF construction involves higher $\Lambda$Ls. However, for neutral excitations of energy two we found that not to be the case.
We have verified all these assertions numerically for small system sizes.

\section{Constructing the ansatz wave functions}
\label{sec:traditional_method}

\subsection{Description}
\label{sec:trad_meth_descr}

From the composite fermion point of view, we can interpret the $\nu=1/2$ Laughlin states in the two-layer description of the bosonic MR ground state (\ref{Pfaff-331}) as Jain ground states corresponding to a $\nu^{*}=1$ IQHE system of CFs (consisting of a boson and one attached vortex). Therefore we propose to construct the excitations over the bosonic MR ground state by creating CF excitations over the Laughlin states in each layer of (\ref{Pfaff-331}). Trial wave functions for excitations over the fermionic MR are obtained by multiplication of the bosonic trial wave functions with an overall Jastrow factor. 

The physical interpretation of this proposal is complicated by the explicit symmetrization in (\ref{Pfaff-331}). A number of works addressing this issue have appeared, focusing on the fermionic case. It was suggested by T.L.~Ho in Ref.~\onlinecite{Ho95} that the spin-polarized MR states can be obtained from the two-layer $331$-state~\cite{Halperin83}, whose spatial wave function does not include the symmetrization, by introducing tunneling between the layers. This proposal turned out to be problematic, and it was argued for example in Refs.~\onlinecite{Read96,Read00,Papic10} that tunneling actually drives the system to an Abelian phase, although at a special value of the parameter that drives the tunneling, a modified version of Ho's model (with some three body interaction added) does describe a critical theory adjacent to the MR phase.  Recently, it was suggested in Ref.~\onlinecite{Papic10} that a weak MR phase can be stabilized around this point, provided that one changes the density of the system at the same time as introducing the tunneling. While this is clearly an important issue, we will not worry too much here about the precise physical mechanism that provides the symmetrization (either for bosons or for fermions), but rather simply write down trial wave functions based on the two layer picture and subject these to numerical scrutiny. 

Explicitly then, we propose to write a general excitation over the bosonic MR ground state as:
\begin{multline}
 \label{Twolayer2}
 \Psi^{exc}_{\rm{Pf}}(z_1,...,z_N) =\\ \mathcal{S} \left(\Psi^{exc,1}_{CF}(z_1,...,z_{N/2}) \times \Psi^{exc,2}_{CF}(z_{N/2+1},...,z_N)  \right)
\end{multline}
where $\Psi^{exc,1}_{CF}$ and $\Psi^{exc,2}_{CF}$ are excitations (neutral, quasihole or quasielectron) of the Laughlin state within the CF picture. For instance, we can create a two-quasielectron state with $z$-angular momentum $l_z$ by taking the $\Psi^{exc,i}_{CF}$ states to be CF quasielectron states with $l_{z_1}$ in layer 1 and $l_{z_2}$ in layer 2 (such that $l_{z_1}+l_{z_2}=l_z$). 

Notice that in Eq.~(\ref{Twolayer2}), the projection onto the LLL is done separately for $\Psi^{exc,1}_{CF}$ and $\Psi^{exc,2}_{CF}$. As mentioned, this construction can be extended to the fermionic cases by multiplying Eq.~(\ref{Twolayer2}) by a global Jastrow factor. This operation being invertible and $L$ preserving, the number of excited states and their angular momentum counting obtained in this way for the fermionic states are the same as those for bosonic states.

It is important to note that obtaining the number of linearly independent angular momentum multiplets is not trivial. As in the case of quasiholes, the trial wave functions for quasielectrons that we propose are not all linearly independent. The LLL projection in each of the layers projects some linear combinations of quasielectron states to zero and the symmetrization between the layers often introduces further linear dependencies. The same considerations apply also to excitons.

We should also point out that the place where the projection onto the LLL is performed is highly relevant. Several others schemes would have been possible. For example, we could have done the projection after the symmetrization. Considering the fermionic states offers even more options since projection can be performed before or after considering the additional global Jastrow factor. Our choice is motivated by the physical picture that the bosonic MR state can be seen as two $\nu = 1/2$ CF layers, both for the ground state and the quasihole excitations.

An alternative set of trial wave functions for quasielectrons and excitons over the fermionic Pfaffian state, presented by Sreejith et al.~in Refs.~\onlinecite{Sreejith11,Sreejith11_PRL_107} can in fact be considered as a differing from the trial wave functions we propose in the way that the LLL-projection is done. Sreejith et al.~define their \emph{bilayer composite fermion} (BCF) wave functions as follows (omitting Gaussian factors),
\begin{equation}
\begin{array}{l}
\label{eq:BCF_exc_Jain}
\Psi^{exc}_{\rm{BCF,\nu=\frac{1}{2}}}(z_1,...,z_N) =\\
\mathcal{A} \left(\Psi^{exc,1}_{\nu=\frac{1}{3}}(z_1,...,z_{N/2}) \times \Psi^{exc,2}_{\nu=\frac{1}{3}}(z_{N/2+1},...,z_N)  \right)\\
\times \prod_{i=1}^{N/2}\prod_{j=N/2+1}^{N} (z_{i}-z_{j}).
\end{array}
\end{equation}
Here the $\mathcal{A}$ stands for total antisymmetrization. As in our own proposal, the electrons have been split into two groups, or layers. However, here each layer is in a $\nu=\frac{1}{3}$ Laughlin (or CF) state, with potentially some quasiholes, quasielectrons or excitons, all created according to the usual CF construction. There is also repulsion between electrons in different layers, but this only induces a single zero in the wave function when two such electrons approach. Finally the wave functions is antisymmetrized in order to make the electrons all indistinguishable. We have taken the number of electrons in each layer equal to $N/2$ in the expression above, but these numbers can in principle be different and must be different if the total number of electrons is odd (as in Ref.~\onlinecite{Sreejith11}). In order to compare these wave functions to our own, we divide by a Jastrow factor (recall our own wave functions for fermions are obtained by multiplying the wave functions (\ref{Twolayer2}) by a Jastrow factor). This leads to the bosonic BCF wave functions 
\begin{equation}
\begin{array}{l}
\label{eq:BCF_exc_Jain_bosonic}
\Psi^{exc}_{\rm{BCF,\nu=1 {\rm ~bosons}}}\,(z_1,...,z_N) =\\
\mathcal{S} \left(\frac{\Psi^{exc,1}_{\nu=\frac{1}{3}}(z_1,...,z_{N/2})}{\prod_{i<j=1}^{N/2} (z_{i}-z_{j})} \times \frac{\Psi^{exc,2}_{\nu=\frac{1}{3}}(z_{N/2+1},...,z_N)}{\prod_{i>j=N/2+1}^{N} (z_{i}-z_{j})}  \right)\\
\end{array}
\end{equation}
This expression obviously leads to the same groundstate and quasihole states as equation (\ref{Twolayer2}). However, for excitons and quasielectrons, where the LLL-projection in the individual layers' wave functions is nontrivial, there are differences, because LLL-projection does not commute with multiplication by a Jastrow factor. Despite these differences, one would expect that the two sets of wave functions have large overlaps, probably describing the same universality class of FQH states. In fact, in numerical studies of systems with a large number of composite fermions, one usually does not use the canonical LLL-projection, but instead the method introduced by Jain and Kamilla in Refs.~\onlinecite{Jain97_IJMPB_22,Jain97_PRB_55}. Similarly, for bosonic systems with a large number of particles, one typically uses the wave functions obtained by first multiplying with a Jastrow factor, then projecting using the Jain-Kamilla method and then dividing out the Jastrow factor again (this practice was introduced in Ref.~\onlinecite{Chang05}). If we were to follow both of these conventions, then the wave functions (\ref{Twolayer2}) and (\ref{eq:BCF_exc_Jain_bosonic}) would become identical. We have in fact done our numerical work using the exact LLL projection in each layer so that we really study different wave functions from those proposed by Sreejith et al. However, as a check, we have also done some calculations of the number of independent states for small system sizes using the alternative LLL-projection which leads to the wave functions of Sreejith et al. This gave essentially the same results as obtained for our own wave functions, but different results from those found by Sreejith et al. More detail on this can be found in Section~\ref{sec:trad_meth_impl}.

\subsection{Vanishing properties of the trial wave functions}
\label{sec:twolayervanishing}

As the vanishing properties of CF wave functions are known (see Section~\ref{sec:CFvanishing}), we can deduce a priori vanishing properties of the two layer states we construct. If $\Psi^{exc,1}_{CF}$ ($\Psi^{exc,2}_{CF}$)  vanishes when $k_1$ ($k_2$) particles are brought to the same point, then $\Psi^{exc}_{\rm{Pf}}$ vanishes when a cluster of $k_1 +k_2 - 1$ particles is formed. For example, a two quasielectron excitation of the MR state can be built from a single quasielectron excitation as depicted in Fig. \ref{fig2}c) in each layer. Such a state is automatically a zero energy state of the Hamiltonian $\mathcal{H}_B^{(5)}$ (see Eq.~\eqref{3bHam}). While such states cannot be a zero energy eigenstates of $\mathcal{H}_B^{(3)}$, since the MR state is the densest zero energy ground state of this Hamiltonian, suitable linear combinations of the quasielectron states  may still be zero energy eigenstates of the $\mathcal{H}_B^{(4)}$ Hamiltonian.

In a similar way, we can deduce the vanishing properties when multiple clusters of electrons are formed in the states constructed by equation (\ref{Twolayer2}) from the cluster vanishing properties of the Laughlin quasielectron states. When $\Psi^{exc,1}_{CF}$ and $\Psi^{exc,2}_{CF}$ are Laughlin states with 1 quasielectron, the resulting states will vanish when a cluster of $4$ particles and a cluster of $3$ particles are formed and when $3$ clusters of $3$ particles are formed. If $\Psi^{exc,1}_{CF}$ and $\Psi^{exc,2}_{CF}$ are Laughlin states with 2 quasielectron states, the resulting states must vanish when a cluster of $4$ particles and a cluster of $5$ particles are formed. This property is trivially satisfied since these states already vanish when any $5$ particles are brought to the same point. However, one may also deduce non trivial vanishing properties: such states vanish when $2$ clusters of $4$ and a cluster of $3$ particles are formed, or when $1$ cluster of $4$ particles and $3$ clusters of $3$ particles are formed, or when $5$ clusters of $3$ particles are formed.

The idea behind the construction \eqref{Twolayer2} is the same as for the CFT construction\cite{Hansson09,Hansson09a}, see Section \ref{sec:CFTqe}: both are inherently two-layer constructions. 
Due to the similarities of the two approaches \emph{in each layer}, we expect that they give equivalent descriptions of the low-energy excitations of the MR ground state. 
In particular, one can show that they obey the same vanishing properties as described above, and they yield identical candidate wave functions for a single exciton as well as for quasielectrons.

\subsection{Implementation of state counting and overlap calculations}
\label{sec:trad_meth_impl}

We use two different methods to generate our trial states and compute overlaps. 

The first method calculates the wave functions (\ref{Twolayer2}) in real space and computes the overlaps using Metropolis integration. 
Real space techniques can be used to compute composite fermion wave functions with over $100$ particles if the LLL projection of Refs.~\onlinecite{Jain97_IJMPB_22,Jain97_PRB_55} is used. However, the symmetrization procedure involves a number of terms which grows factorially with the number of particles $N$ and this limits the reachable size with this method to $N = 16$. 

In our second method, we first compute Laughlin states $\Psi^{exc}_{CF}$ with the desired excitations  in the CF picture using the exact method for computing CF wave function explained in Ref.~\onlinecite{2011arXiv1105.5907S}. Then the states are symmetrized at the Fock space level. Using this method, we were able to generate bosonic states up to $20$ particles. Using Schur polynomials that can be generated using recursion formulas from Ref.~\onlinecite{PhysRevLett.103.206801}, we can multiply these wave functions by a global Jastrow factor at the Fock space level, converting our bosonic states to fermionic ones. This can be done for up to $14$ particles (Hilbert spaces of a few hundred thousands of independent states). 

A major advantage of the Fock space method over the real space method is that all calculations are done at the machine precision and results are expressed in the $n$-body basis. In the real space method, overlap calculations are done by Monte Carlo integration and suffer from statistical errors which are much larger than machine rounding errors. A potential advantage of the real space method is that it makes multiplication by a Jastrow factor trivial, which means fermionic calculations could in principle be done for up to $16$ particles using this method. 

We now describe our calculation of the number of linearly independent trial wave functions at each value of $L_z$. Formula (\ref{Twolayer2}) provides, for each value of $L_z$, and for each number of flux quanta and quasielectrons or excitons, a set of wave functions $\Psi_i, i=1,\ldots,d$. These wave functions are usually not linearly independent as both symmetrization and projection induce linear dependencies. We want to find a basis of linearly independent states $\chi_{i},i=1,\ldots,d'\le d$ for the space spanned by the $\Psi_{i}$. To do this, we compute the overlap matrix, given by
\ba
M_{ij} &=&\braket{\Psi_i}{\Psi_j},
\label{limatrix}
\ea
and we diagonalize it. In the ideal case, the matrix $M$ has a number of eigenvalues which are clearly non-zero and a number of eigenvalues which equal zero to numerical accuracy. This actually occurs in all our calculations when we use the Fock space method. Since $M$ is hermitian, there exists a unitary matrix $Q$ such that $QMQ^{-1}$ is diagonal, that is, $\sum_{j,k}Q_{ij}M_{jk}Q^{-1}_{kl}=\lambda_{i}\delta_{il}$, where $\lambda_{i}$ are the eigenvalues of $M$. We now define $\chi_{i}=\sum_{j}\bar{Q}_{ij}\Psi_{j}$ and it follows that $\braket{\chi_{i}}{\chi_{j}}=\delta_{i,j}\lambda_{i}$. This implies that whenever $\lambda_{i}=0$, we also have $\chi_{i}=0$, giving a linear relation between the $\Psi_{i}$. The $\chi_{i}$ which belong to nonzero eigenvalues, $\lambda_{i}\neq 0$ form the sought after orthogonal basis for the vector space spanned by the $\Psi_{i}$. Hence, the dimension of the space spanned by the trial wave functions $\Psi_i$ is just the number of nonzero eigenvalues of the overlap matrix $M$. 

The method described above works very well for us when we use the Fock space method of evaluating the wave functions and overlaps. However, if we use the real space method to calculate the overlap matrix, there are statistical errors in the matrix which cause its spectrum to have a number of spurious small but nonzero eigenvalues, which should be discarded to obtain the correct counting of states. Moreover, it is important to realize that the dimension of the space spanned by a set of trial wave functions is not a quantity that is stable under perturbation; random perturbation of a set of linearly dependent wave functions $\Psi_i$ will tend to make them all linearly independent. Such perturbations may be introduced for example by making a change in the way that lowest Landau projection is implemented. In such cases, as long as the spurious eigenvalues of $M$ are small enough, one may approximate the original trial wave functions $\Psi_{i}$ very well using only those vectors $\chi_{i}$ which correspond to the larger eigenvalues. Explicitly, let us first define the orthonormal basis vectors $\tilde{\chi}_{i}=\frac{\chi_{i}}{\sqrt{\lambda_{i}}}$.  We may expand each of the $\Psi_{i}$ in terms of these, that is, we may write $\Psi_{i}=\sum_{j}c_{ij}\tilde{\chi}_{j}$ for some coefficients $c_{ij}$. We then find that $c_{ij}=\braket{\tilde{\chi}_{j}}{\Psi_{i}}=\sqrt{\lambda_{j}} Q_{ji}$. Since $Q$ is a unitary matrix, $|Q_{ij}|\le 1$ and hence $|c_{i,j}|^2\le \lambda_{j}$. Hence we see that it is an excellent approximation, in terms of the quantum mechanical inner product, to drop the states corresponding to the small eigenvalues of $M$ from our description of the space spanned by the $\Psi_{i}$, as long as the sum of the $M$-eigenvalues of the dropped states is much smaller than $1$. In practice we find in our calculations using the real space method that it is always possible to make a cut in the spectrum of $M$ which satisfies this condition very well, and when this is done we obtain the same counting of states as that obtained using the Fock space method. 

Once we have a basis of linearly independent states $\tilde{\chi}_i, i=1,...\tilde{d}$ for the Hilbert space $\mathbb{H}^{L_z}$ we can compute overlaps between this set of wave functions and another set of wave functions obtained, for example, by exact diagonalization. To do this, we need to generalize the notion of overlap between single states to overlap between subspaces. If we have two bases of normalized states $\phi_i, i=1,..,d$ and $\eta_i, i=1,...,d$ and we are interested to know if they generate the same subspace, we can take the trace of the operator that projects one of the bases into the other, i.e.~we can define the (squared) overlap between the two subspaces as:
\ba
Overlap = \frac{1}{d} \sum_{i=1}^{d} \sum_{j=1}^{d} \mid< \phi_i| \eta_j>\mid^2 \ . \nl
\label{overlap}
\ea
This quantity (\ref{overlap}) is a natural measure for the overlap between two different subspaces. In particular, if the subspace generated by both bases is the same, the overlap is equal to one, and if they don't generate the same subspace it is easy to see that (\ref{overlap}) is less than one.

\section {NUMERICAL RESULTS}
\label{sec:numerical_results}

In this section, we present the results of our numerical calculations. In Section~\ref{sec:num_result_counting} we show that our ansatz wave functions produce, for the boundary excitations on the disc, the same counting formulas for quasiholes and quasielectrons. We also conjecture a counting formula (for any value of $L_z$) for two and four-quasielectron states. In Section~\ref{sec:BHcomp}, we then  compute the root partitions (see Section~\ref{sec:Jacks}) of our quasielectron states and we show that our states share a number of properties with the quasielectrons obtained by Bernevig and Haldane in Ref.~\onlinecite{Bernevig09_PRL_102}. Section~\ref{sec:unbalanced} contains a discussion on the relevance of unbalanced states, which have different number of electrons in the two layers. Finally, in Section~\ref{sec:num_result_spectra} we show how well our construction describes the low energy spectrum of both model and realistic Hamiltonians. We also show that the subset of our trial states which vanish when a cluster of $4$ electrons is formed is particularly successful in attaining low variational energies and large overlaps with exact wave functions.

\subsection{Multiplet counting for the trial wave functions} 
\label{sec:num_result_counting}

As discussed in Section~\ref{sec:trad_meth_descr} both the projection and the symmetrization can create linear dependencies between the ansatz wave 
functions (\ref{Twolayer2}). We have computed, for each $L_z$, and for different numbers of particles the dimension of the space of linearly independent trial wave functions for the various types of excitations. The calculations have been performed up to $N=20$ involving Hilbert spaces as large as $6\times 10^7$. For quasielectrons at high values of $L_z$, these dimensions match perfectly with the values predicted by the CFT describing the boundary excitations (low energy excitations) of the MR Pfaffian phase on the disc (see Section~\ref{sec:CFTreview}). Therefore our ansatz wave functions for quasiparticles show the same topological properties for quasielectrons as for quasiholes.

In Tables \ref{tab:qpcounting}-\ref{tab:exccounting} we give the numbers of independent states we have found for various numbers quasielectrons and excitons, for each value of the total angular momentum $L$ and for different numbers of electrons.

\begin{widetext}
\begin{center}
\begin{table}[htb]
\begin{center}
\begin{tabular}{ |r| l | c | c |c |c |c |c |c |c |c |c |c |c |c |c |c |c | c | c | r |}
   \hline
   &\bf{N} / \bf{L}  & 0 & 1 & 2 & 3 & 4 &  5  & 6 & 7 & 8  & 9 & 10 & 11 & 12 & 13 & 14 & 15 & 16 & 17 & 18\\ \hline
  $n=2$&\hspace{0.25cm}  14   & 0 & 1 & 0 & 1 & 0 & 1 & 0 & 1 & - & - & - & - & - & - & - & -& - & - & -\\ 
  &\hspace{0.25cm}  16  & 1 & 0 & 1 & 0 & 1 & 0 & 1 & 0 & 1 & - & - & - & - & - & - & - & - & - & -\\  
  &\hspace{0.25cm}  18  & 0 & 1 & 0 & 1 & 0 & 1 & 0 & 1 & 0 & 1 & - & - & - & - & - & - & - & - & -\\ \hline
  $n=4$&\hspace{0.25cm}  14   & 2 & 0 & 4 & 1 & 4 & \bf{2} & \bf{3} & \bf{1} & \bf{2} & \bf{0} & \bf{1} & - & - & - & - & -& - & - & -\\ 
  &\hspace{0.25cm}  16   & 3 & 0 & 4 & 2 & 5 & 2 & \bf{5} & \bf{2} & \bf{3} & \bf{1} & \bf{2} & \bf{0} & \bf{1} & - & - & -& - & - & -\\ 
  &\hspace{0.25cm}  18   & 3 & 0 & 5 & 2 & 6 & 3 & 6 & \bf{3} & \bf{5} & \bf{2} & \bf{3} & \bf{1} & \bf{2} & \bf{0} & \bf{1} & -& - & - & -\\ \hline
$n=6$&\hspace{0.25cm}  14   & 0 & 3 &  1 & 5 & 2 & 3 & \bf{2} & \bf{2} & \bf{0} & \bf{1} & - & - & - & - & - & - & - & - & -\\ 
  &\hspace{0.25cm}  16   & 3 & 1 & 6 & 4 & 8 & 4 & 7 & 3 & \bf{4} & \bf{2} & \bf{2} & \bf{0} & \bf{1} & - & - & - & - & - & -\\ 
  &\hspace{0.25cm}  18   & 0 & 7 & 4 & 11 & 8 & 12 & 9 & 11 & 6 & 8 & \bf{4} & \bf{4} & \bf{2} & \bf{2} & \bf{0} & \bf{1} & - & - & -\\ 
  &\hspace{0.25cm}  20   & na & na & na & na & na & na & na & na & na & na & na & 7 & \bf{8} & \bf{4} & \bf{4} & \bf{2} & \bf{2} & \bf{0} & \bf{1} \\ \hline
  $n=8$&\hspace{0.25cm}  14   & 1 & 0 & 1 & \bf{0} & \bf{1} & - & - & - & - & - & - & - & - & - & - & -& - & - & -\\ 
  &\hspace{0.25cm}  16   & 2 & 0 & 3 & 1 & 3 & 1 & \bf{2} & \bf{0} & \bf{1} & - & - & - & - & - & - & - & - & - & -\\ 
  &\hspace{0.25cm}  18   & 4 & 1 & 6 & 4 & 8 & 4 & 7 & 3 & 4 & \bf{2} & \bf{2} & \bf{0} & \bf{1} & - & - & - & - & - & - \\ \hline
\end{tabular}
\caption{Multiplicities of the angular momentum multiplets for $n$-quasielectron states. The dash symbol indicates that there is no state and na stands for non available. The dimensions corresponding to the highest values of $L$ in boldface stabilize when we increase the number of particles and matches the dimensions given by the CFT describing the boundary excitations on the disc. }
\label{tab:qpcounting}
\end{center}
\end{table}
\end{center}
\end{widetext}
Table~\ref{tab:qpcounting} shows the numbers of independent multiplets of our trial states with $2$ to $8$ quasielectrons, for $N=14$, $N=16$ and $N=18$. 
Note that the dimensions corresponding to the higher values of $L$ (in boldface) stabilize when we increase the number of particles. As explained in Section~\ref{sec:CFTreview}, they correspond to the dimensions given by the CFT describing the boundary excitations on the disc. 
Using Eq. (\ref{qhole-count}), one may check that the stable multiplicities in Table \ref{tab:qpcounting}, are the same as those observed in the high $L$ sector of systems with $2$ to $8$ quasiholes.
\begin{center}
\begin{table}[htb]
\begin{tabular}{| l | c | c |c |c |c |c |c |c |c |c |c |c |c |c |c | c| c| r |}
   \hline
   \bf{N} / \bf{L}  & 0 & 1 & 2 & 3 & 4 &  5  & 6 & 7 & 8  & 9 & 10 & 11 & 12 & 13 & 14 & 15 & 16\\ \hline
   \hspace{0.25cm}  12   & 2 & 3 & 6 & 7 & 8 & 6 & \bf{7} & \bf{5} & \bf{3} & \bf{2} & \bf{1} & - & - & - &  - & - & -  \\
      \hspace{0.25cm}  14   & 1 & 5 & 7 & 10 & 10 & 11 & 9 & \bf{9} & \bf{7} & \bf{5} & \bf{3} & \bf{2} & \bf{1} & - & - & - &  -    \\
   \hspace{0.25cm}  16   & 2 & 5 & 10 & 11 & 14 & 14 & 14 & 12 & \bf{12} & \bf{9} & \bf{7} & \bf{5} & \bf{3} & \bf{2} & \bf{1} & - &  - \\ 
      \hline 
   \hspace{0.25cm}  12   & 2 & 3 & 6 & 7 & 9 & 9 & 10 & 8 & \bf{7} & \bf{5} & \bf{3} & \bf{2} & \bf{1} & - & - & - &  -   \\
      \hspace{0.25cm}  14   & 1 & 5 & 7 & 10 & 11 & 13 & 12 & 13 & 11 & \bf{9} & \bf{7} & \bf{5} & \bf{3} & \bf{2} & \bf{1} & - &  -  \\
   \hspace{0.25cm}  16   & 3 & 5 & 9 & 12 & 15 & 15 & 17 & 16 & 16 & 14 & \bf{12} & \bf{9} & \bf{7} & \bf{5} & \bf{3} & \bf{2} & \bf{1} \\ \hline
\end{tabular}
\caption{Multiplicities of the angular momentum multiplets for: a system with one exciton (in the second $\Lambda$L) over a two-quasielectron ground 
state (top) and a system with one exciton (in the second $\Lambda$L) over a two-quasihole groundstate (bottom).}
\label{tab:xcandqpcounting}
\end{table}
\end{center}
The number of multiplets of single exciton states is just the same as the number of single exciton states over one of the two layers of composite fermions. This was to be expected as the trial wave functions for single excitons have an exciton in one composite fermion layer and a ground state wave function in the other. We find that a system of $N$ particles has no multiplets at $L=0$ and $L=1$ and has a single multiplet at all subsequent $L$-values up to the maximal $L$-value where a multiplet occurs, which is $L=N/2$.

Table~\ref{tab:xcandqpcounting} shows the multiplicities corresponding to states with a single exciton and two quasielectrons (top) and a single exciton and  two quasiholes (bottom). As in the case of quasielectrons and quasiholes without excitons present, we obtain the same numbers for quasielectrons and quasiholes at high $L$, once again confirming the idea that the CFT and TQFT describing the quasielectrons should be the same as for the quasiholes. We may also conjecture a formula for the bold numbers in these tables. They equal the integer parts of $\frac{(p+1)(p+2)}{6}$, where $p$ is the position of the $L$-value counting from the highest $L$ for which a multiplet of trial states exists. Alternatively, we may characterize these numbers by the generating function $\frac{x}{(1-x)^2(1-x^3)}$. 
\begin{center}
\begin{table}[htb]
\begin{tabular}{ | l | c | c |c |c |c |c |c |c |c |c |c |c |c |c |c |c | c | r |}
   \hline
   \bf{N} / \bf{L}  & 0 & 1 & 2 & 3 & 4 &  5  & 6 & 7 & 8  & 9 & 10 & 11 & 12 & 13 & 14 & 15 & 16 \\ \hline
    12   & 5 & 3 & 10 & 7 & 13 & 9 & 13 & 8 & \bf{9} & \bf{4} & \bf{4} & \bf{1} & \bf{1} & - & - & - & -    \\
    14   & 6 & 4 & 13 & 10 & 18 & 13 & 18 & 13 & 15 & \bf{9} & \bf{9} & \bf{4} & \bf{4} & \bf{1} & \bf{1} & - & -    \\
    16  & 7 & 5 & 16 & 13 & 23 & 18 & 24 & 18 & 22 & 15 & \bf{16} & \bf{9} & \bf{9} & \bf{4} & \bf{4} & \bf{1} & \bf{1} \\ \hline 
\end{tabular}
\caption{Multiplicities of the angular momentum multiplets for a system containing excitons with effective CF energy 2.}
\label{tab:exccounting}
\end{table}
\end{center}
Table~\ref{tab:exccounting} shows the multiplicities corresponding to a system with up to two excitons. Observe in the table that the stable multiplicities occur in pairs and that they are simply the squares of the integers ($1$, $4$, $9$, $16$ etc.). Double excitons can in principle occur in three different ways. One may excite a particle to the first unoccupied $\Lambda$L in each of the composite fermion layers, or one may excite two particles in a single layer, or one may excite a single particle up to the second unoccupied $\Lambda$L. We have considered all of these possibilities, as they all occur at the same naive composite fermion energy. However, it is worth noting that, to obtain the entire set of linearly independent trial wave functions, one does not actually need to include the states with two excitons in the same layer or with particles excited to the second unoccupied $\Lambda$L. We checked numerically that the states constructed in this way are already contained in the space of trial wave functions with a single exciton in each layer.

We have already conjectured counting formulae for the stable numbers of multiplets of quasielectron and exciton states at high $L$, relating quasihole and quasielectron states. However, for $2$ and $4$-quasielectron states, we may go further and conjecture counting formulae for any $L$-value.  
Observe that, \emph{before projection and symmetrization,} the number of independent $n$-quasielectron states (considering second $\Lambda$L quasielectrons only) in a system with $N$ particles, is the same as the number of independent $n$-quasihole states in a system with $N'$ particles, where $N'$ is given by:
\ba
N'=N - 2*n + 4 \ .
\label{qh-qe-duality}
\ea
Of course the naive counting of states is modified as the projection and symmetrization operations introduce linear dependencies between the ansatz wave functions. Nevertheless, for all cases we have checked, it turns out that after projection and symmetrization, these countings, though modified by projection and symmetrization, are still equal for quasiholes and quasielectrons, for all $L$, in the particular cases of $2$ and $4$-quasihole/quasielectron states (we checked this up to $N=18$). We conjecture that this equality holds for all $N$.  For higher numbers of quasiholes/quasielectrons, the countings are no longer the same at all $L$-values. However, by inspection we note that even in these cases, equality of multiplicities between quasihole and quasielectron states still holds for some $L$-values beyond the stable ones.

Some further relations between multiplet countings at different numbers of particles and different numbers of electrons can be conjectured using particle hole duality in the composite fermion $\Lambda$-levels. One may naively conjecture that the same counting should be obtained for a system with $N$ particles and $n$ quasielectrons as for a system with $2N-3n+4$ particles and $2(N/2-n+2)$ quasielectrons (this is true before LLL-projection and symmetrization). We observe in our data that this holds for $n=2$ and $n=4$, but not beyond $4$ quasielectrons.

\subsection{Vanishing properties and comparison to Bernevig-Haldane construction}
\label{sec:BHcomp}

Now we come back to the Bernevig-Haldane quasielectron and exciton states introduced in Section~\ref{sec:BH} and we will compare them with our ansatz wave functions. The easiest comparison to make is between our $1$-exciton states and those of Bernevig and Haldane which have a single non-Abelian quasielectron and a single non-Abelian quasihole. We have checked that our single exciton states have the same root partitions as the Bernevig-Haldane excitons (given explicitly in formula~(\ref{eq:rootNA})). We also find that they satisfy the same vanishing properties (\ref{eq:clusteringNA}) and (\ref{eq:extravanish}). Hence, for single excitons, our trial wave functions are in fact the same as those proposed by Bernevig and Haldane. 

For wave functions with multiple excitons or with only quasielectrons, it is a bit more complicated to make a comparison between our trial wave functions and the ones proposed by Bernevig and Haldane, if only because the root partitions for such states are not given explicitly in Ref.~\onlinecite{Bernevig06}. However, it is clear that, in these more general cases there can be some mismatch between the two constuctions. For example, let us compare our construction to Bernevig and Haldane's construction for non-Abelian quasielectrons. 

As discussed in Section~\ref{sec:twolayervanishing}, our two-quasielectron states vanish when $5$ particles cluster together. They also vanish when $3$ clusters of $3$ particles are formed and when $2$ clusters, one of $4$ particles and one of $3$ particles are formed. These vanishing properties are also satisfied by our $2$-exciton states. However, the non-Abelian quasielectron states of Ref.~\onlinecite{Bernevig06} vanish already when a single cluster of $4$ particles is formed and also when $2$ clusters of $3$ particles are formed. It is easy to check directly that not all of our trial wave functions satisfy these stronger vanishing properties, so we obtain a mismatch with Bernevig and Haldane's construction. 

It is interesting to look at subspaces of our space of trial wave functions which do satisfy stronger vanishing properties, such as those required by Bernevig and Haldane. One motivation for this is that it may be (naively) expected that trial wave functions which vanish already when $4$ particles cluster together may have lower variational energy (for repulsive potentials) than wave functions which don't vanish until a cluster of $5$ particles is formed. This indeed turns out to be the case (see Section~\ref{sec:num_result_spectra}). We have therefore studied in some detail the subspaces of our spaces of trial wave functions which consists of those states which vanish when a cluster of $4$ particles is formed. In other words, these are the space of zero modes of the $\mathcal{H}_{B}^{(4)}$ Hamiltonian inside our spaces of trial wave functions. We give a sample of the results for the counting of such states in Table~\ref{tab:HB4qpcounting}. By comparison to Table~\ref{tab:qpcounting} we see immediately that, indeed, not all our trial states are zero modes of $\mathcal{H}_{B}^{(4)}$. In particular, the trial states at the highest angular momenta seem never to vanish when $4$ particles cluster together.
\begin{center}
\begin{table}[htb]
\begin{tabular}{ |r| l | c | c |c |c |c |c |c |c |c |c |c |c |c | c | c | r |}
   \hline
   &\bf{N} / \bf{L}  & 0 & 1 & 2 & 3 & 4 &  5  & 6 & 7 & 8 \\ \hline
  $n=2$&\hspace{0.25cm}  14   & 0 & 1 & 0 & 1 & 0 & 1 & - & - & - \\ 
  &\hspace{0.25cm}  16        & 1 & 0 & 1 & 0 & 1 & 0 & 1 & - & - \\  
  \hline
  $n=4$&\hspace{0.25cm}  14   & 1 & 0 & 2 & 0 & 1 & - & - & - & -\\ 
  &\hspace{0.25cm}  16        & 2 & 0 & 2 & 1 & 2 & 0 & 1 & - & -\\
  \hline
\end{tabular}
\caption{Multiplicities of angular momentum multiplets for $n$-quasielectron states which are also zero modes of $\mathcal{H}_{B}^{(4)}$.}
\label{tab:HB4qpcounting}
\end{table}
\end{center}
We find similar results to those presented in Table~\ref{tab:HB4qpcounting} for the counting of multiplets of $2$-exciton states which are also zero modes of $\mathcal{H}_{B}^{(4)}$. Again, not all our trial states for double excitons are zero modes of $\mathcal{H}_{B}^{(4)}$ and in particular, the trial states at the two highest angular momenta never vanish when $4$ particles cluster together. For the case of double excitons, we can give an exact description of the numerical results for multiplet counting by a formula which relates the number of multiplets which are zero modes of $\mathcal{H}_{B}^{(4)}$ to the total number of multiplets. We observe that the total number of multiplets equals the number of multiplets of $\mathcal{H}_{B}^{(4)}$ zero modes for $L=0,1,2,3$  and then from $L=4$ upward, the number of multiplets of $\mathcal{H}_{B}^{(4)}$ zero modes is lower than the total number of multiplets, by $\min\{[L/2-1],[(N-L)/2+1]\}$, where $\min$ denotes the minimum and the square brackets denote the integer part. 

We also calculated the root partitions for the two-quasielectron states which are also zero modes of $\mathcal{H}_{B}^{(4)}$. For $N=16$, these are given in Table~\ref{tab:2qeroot}. These root configurations correspond to the ones predicted by Bernevig and Haldane's construction~\cite{bernevigPC} for states with $2$ non Abelian quasielectrons.
\begin{table}[htb]
\begin{tabular}{|c|c|}
\hline
$L_z$ & highest root configuration\\ \hline
0 & 3 0 1 1 1 1 1 1 1 1 1 1 0 3 \\ \hline
1 & 2 1 1 1 1 1 1 1 1 1 1 1 0 3 \\ \hline
2 & 2 0 2 1 1 1 1 1 1 1 1 1 0 3 \\ \hline
3 &2 0 2 0 2 1 1 1 1 1 1 1 0 3 \\ \hline
4 &2 0 2 0 2 0 2 1 1 1 1 1 0 3 \\ \hline
5 &2 0 2 0 2 0 2 0 2 1 1 1 0 3 \\ \hline
6 &2 0 2 0 2 0 2 0 2 0 2 1 0 3 \\ \hline
\end{tabular}
\caption{ 
Highest root configurations of the zero energy states of $\mathcal{H}_{B}^{(4)}$ in each $L_z$ sector of our space of $2$-quasielectron trial states at $N = 16$}
\label{tab:2qeroot}
\end{table}

\subsection{Unbalanced states}
\label{sec:unbalanced}

So far, we have considered only balanced states, that is, states with the same number of electrons in both CF layers. However, there is no strong reason for this restriction a priori. In fact for odd numbers of electrons, unbalanced states with an odd number of electrons in one layer and an even number in the other layer are unavoidable, and as shown in Ref.~\onlinecite{Sreejith11} these indeed give a good description of the lowest band of states for an odd number of electrons in the MR Pfaffian phase. In this paper, we are looking at systems where the total number of electrons is even, and in this case, we find that, at least in the low energy sector of the theory (defined with respect to the CF energy), unbalanced states don't introduce new physics and can be ignored. The rest of this section is devoted to explaining this in some detail and can be skipped on first reading.

We will deal with unbalanced quasihole states in the bosonic case, but the same results hold for fermions. 
Consider a 2n-quasihole state with $N=N_1+N_2$ particles where $N_1$ and $N_2$ are the 
number of particles in the first and second layers respectively and suppose that $N_1>N_2$. 
We will require that both layers feel the same magnetic field. This requirement can be physically motivated by the idea that the Pfaffian phase can really be viewed as some kind of modified two-layer system. The largest exponent of the coordinates in the first layer, $N_{\phi 1}$, must then be equal to the corresponding exponent $N_{\phi 2}$ in the second layer. Note also that these two exponents must equal the total number of fluxes in the system, $N_\phi=N - 2 + n $. Hence, we have
\ba
 N_{\phi 1} &=& N - 2 + n = 2 N_1 - 2 + (N_2 - N_1 + n) \nl
 N_{\phi 2} &=& N - 2 + n = 2 N_2 - 2 + (N_1 - N_2 + n).
\label{fluxes}
\ea
From this equation, it is clear that the second layer corresponds to a $\nu=1/2$ Laughlin state with $N_1 - N_2 + n$ quasiholes. For the first 
layer we have two different cases (depending on the values of $N_1$ and $N_2$) corresponding to a $\nu=1/2$ Laughlin state with either $N_1 - N_2 + n$ quasiholes 
(if $N_2 - N_1 + n>0$ ) or $N_1 - N_2 + n$ quasielectrons (if $N_2 - N_1 + n<0$). 
In the first case the two layers of the trial wave function both contain $\nu=1/2$ Laughlin quasihole wave functions. These $2n$-quasihole states have the property that they vanish when three particles are at the same position\ (see Section~\ref{sec:BH}). Therefore because the balanced $2n$-quasiholes already span the complete set of wave functions with this vanishing property we can conclude that the unbalanced quasiholes are linear combinations of the balanced ones and we don't need to consider them in our construction. 

The case where one layer has quasiholes and the other quasielectrons is more interesting. 
In particular, in cases with a single quasielectron in one of the layers, these unbalanced states have the property that they vanish whenever a cluster of $4$ particles is formed. This vanishing property is different from the usual vanishing property for balanced quasihole states, which already vanish when $3$ particles form a cluster.  We have verified that these unbalanced states are linearly independent of the balanced wave functions and therefore they are new quasihole states. However, we have two reasons to believe that these unbalanced quasiholes are not relevant to the lowest energy sector of quasihole states. First of all, we can use the composite fermions' kinetic energy, read off from the number of particles in each $\Lambda$L, as a guide.  
The new states include higher $\Lambda$L contributions, while the balanced quasiholes have composite fermions only in the lowest $\Lambda$L, and so, at least naively, we can focus on balanced quasiholes in any low energy description. Secondly, the fact that the new quasihole wave functions vanish only when $4$ particle positions coincide will probably mean they have higher variational energies for realistic repulsive interactions when compared to balanced quasihole states which vanish already when $3$ particle positions coincide.

For unbalanced quasielectron states it can be shown, in a similar way as we did for unbalanced quasiholes, that depending on the values of $N_1$ and $N_2$, we can have either quasiholes or quasielectrons in the second layer, while in the first layer we always have quasielectrons. 
It is easy to see that, in contrast to the situation for unbalanced quasiholes, the unbalanced quasielectrons can, for suitable values of $N_1$, $N_2$ and $n$, produce states with the same $\Lambda$L energy as the balanced ones. However, it is also easy to see, that this only starts to occur when the overall number of quasielectrons is at least $4$. In the $4$-quasielectron sector we have compared balanced and unbalanced $4$-quasielectron states with the same $\Lambda$L energy and we have found that the unbalanced states are linear combinations of the balanced ones. Therefore they can be omitted from our construction. We have similarly investigated a number of other cases where unbalanced quasielectron states with the same $\Lambda$L energy as the balanced quasielectron states exist, and in all cases considered, the unbalanced states were already contained in the space spanned by the balanced states. 

Finally for the exciton sector (where $n=0$) it is easy to see that unbalanced excitons correspond to states with $N_1-N_2$ quasielectrons in the first layer and with the same number of quasiholes in the second layer, in addition to any excitons that may exist within the layers (this was also noted in Ref.~\onlinecite{Sreejith11}).
It can be shown that, also in this case, for certain values of $N_1$ and $N_2$ it is possible to construct low energy unbalanced states with the same $\Lambda$ level energy as the balanced excitons studied before. For single excitons this does not happen, as long as $N$ is even, but for double excitons, with $\Lambda$L energy $2$, there are unbalanced states at the same $\Lambda$L energy with $N_1-N_2=2$: $2$ quasielectrons in the first layer and $2$ quasiholes in the second layer. 
However, we checked that these unbalanced excitons are linear combinations of the balanced two-exciton states and once again we do not need to include them in our construction.

\subsection{Comparison of Spectra and Overlaps}
\label{sec:num_result_spectra}

To test our construction, we now compare the exact spectra of realistic and model Hamiltonians in the LLL with the spectra of these same Hamiltonians in the spaces spanned by our trial wave functions. We do this both for bosons and for fermions. For the bosonic states, we use the $\mathcal{H}_B^{(3)}$ Hamiltonian, for which the MR state and its quasihole states are zero-energy eigenstates, as our model Hamiltonian. We use the $\mathcal{H}_B^{(2)}$ Hamiltonian as our realistic Hamiltonian. This is well justified, since in most experiments with ultra-cold bosonic gases, the relevant interaction is s-wave scattering \cite{2008AdPhy..57..539C}, which is modeled well by this potential. 
For the fermionic states, the realistic interaction is the Coulomb interaction in the second Landau level. We considered this interaction with the first relevant pseudopotential $V_1$ (which describes the shortest range part of the interaction \cite{Haldane83}) shifted by $ \delta V_1 =0.035$, so that the overlap between the MR ground state and the exact ground state is maximal. Since the fermionic MR state and its quasiholes states are zero-energy eigenvectors of $\mathcal{H}_F^{(3)}$ the 3-body hollow core interaction \cite{PhysRevLett.66.3205} (the analogue of $\mathcal{H}_{B}^{(3)}$ for fermionic systems), we take this as our model Hamiltonian for the fermionic states. Studies of spectra and overlaps comparable to the one presented here can be found in Ref.~\onlinecite{Regnault07} for the bosonic quasiholes states and in Ref.~\onlinecite{Toke2007504} for the fermionic ones. A recent paper by Sreejith et al, Ref.~\onlinecite{Sreejith11_PRL_107}, also presents a number of results on spectra and overlaps closely related to those presented here, for a slightly different family of wave functions (see Section~\ref{sec:trad_meth_descr} for details).

\begin{figure*}[htb]
\begin{center}
\vspace*{-1mm}
\includegraphics[width= 8cm]{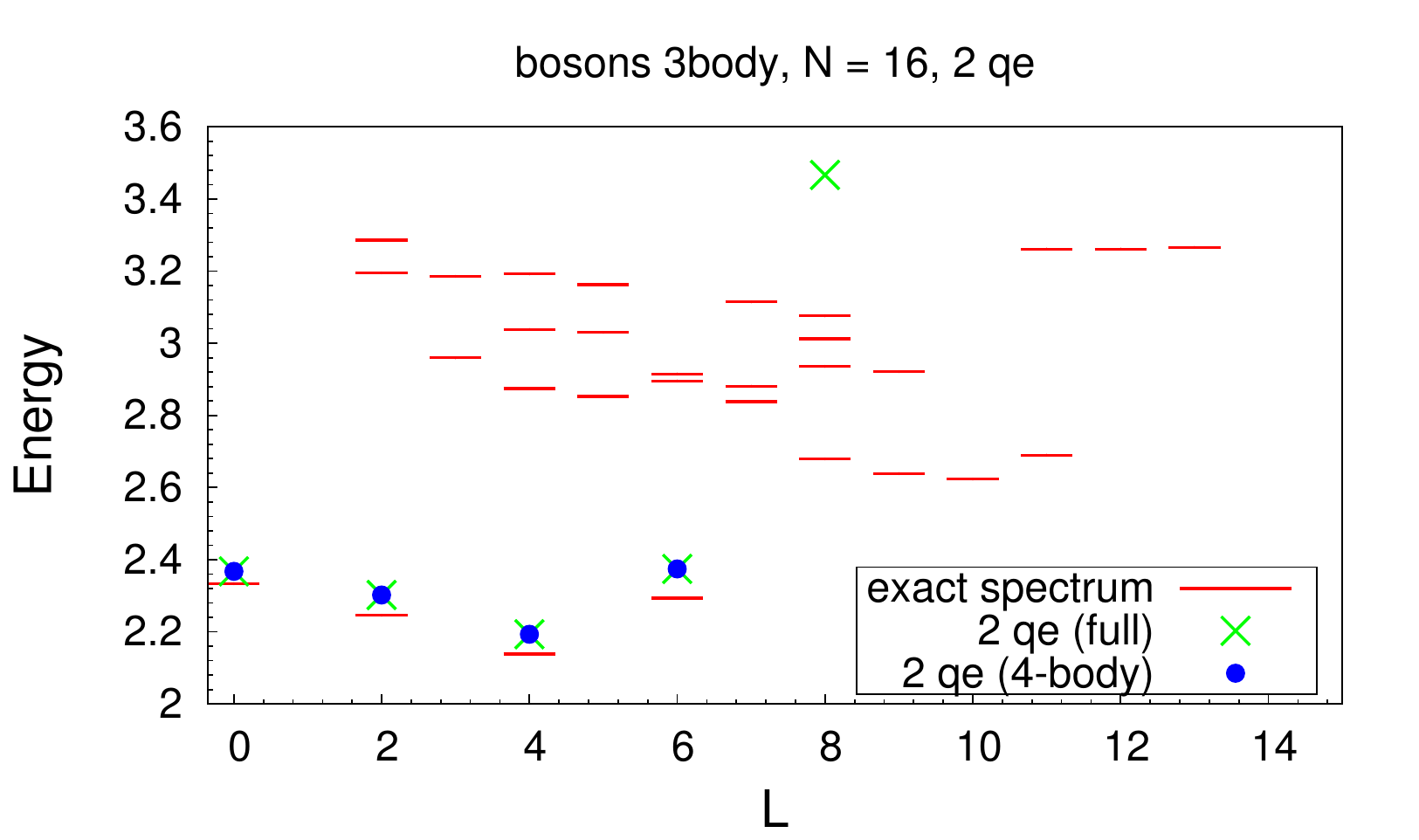}
\includegraphics[width= 8cm]{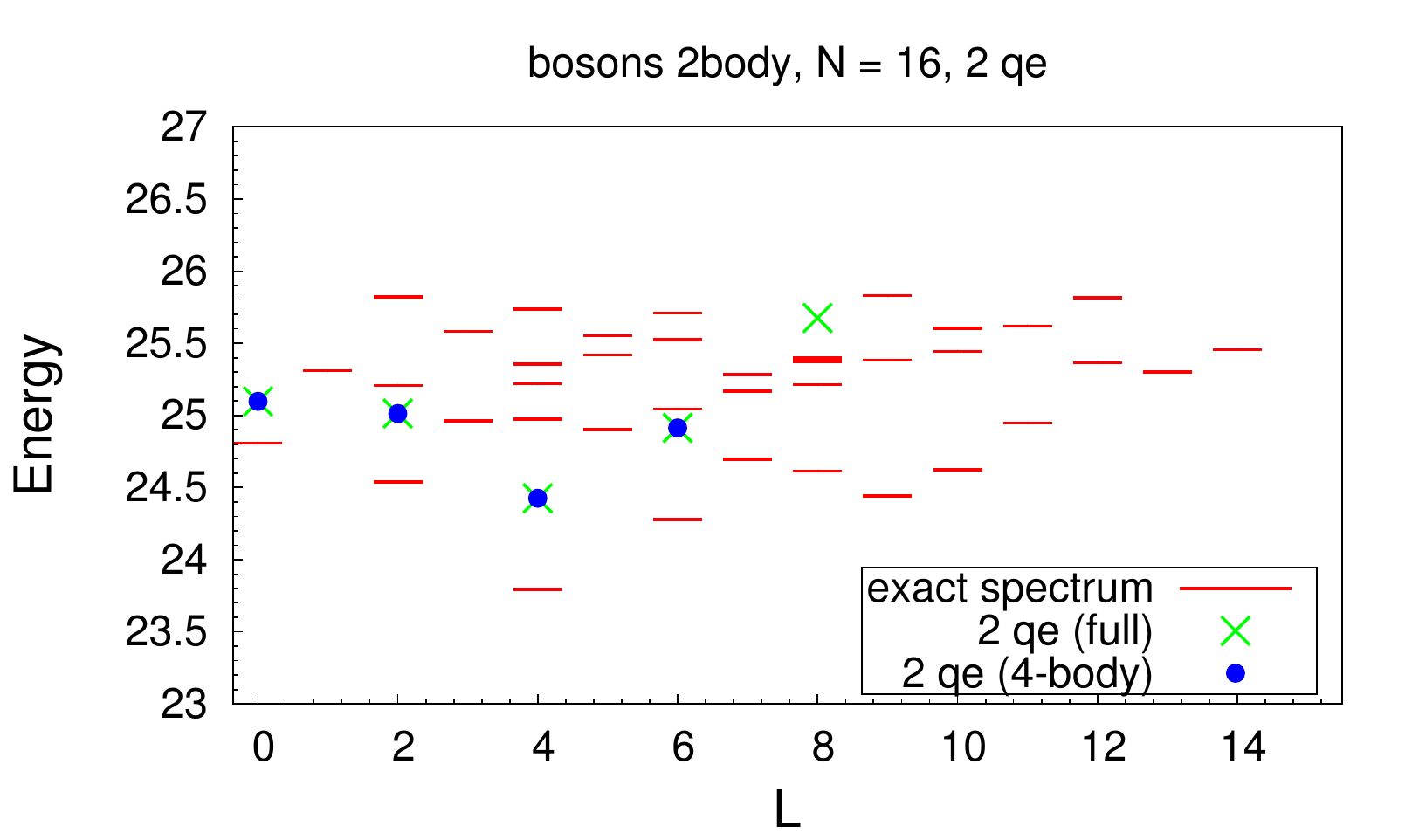}
\includegraphics[width= 8cm]{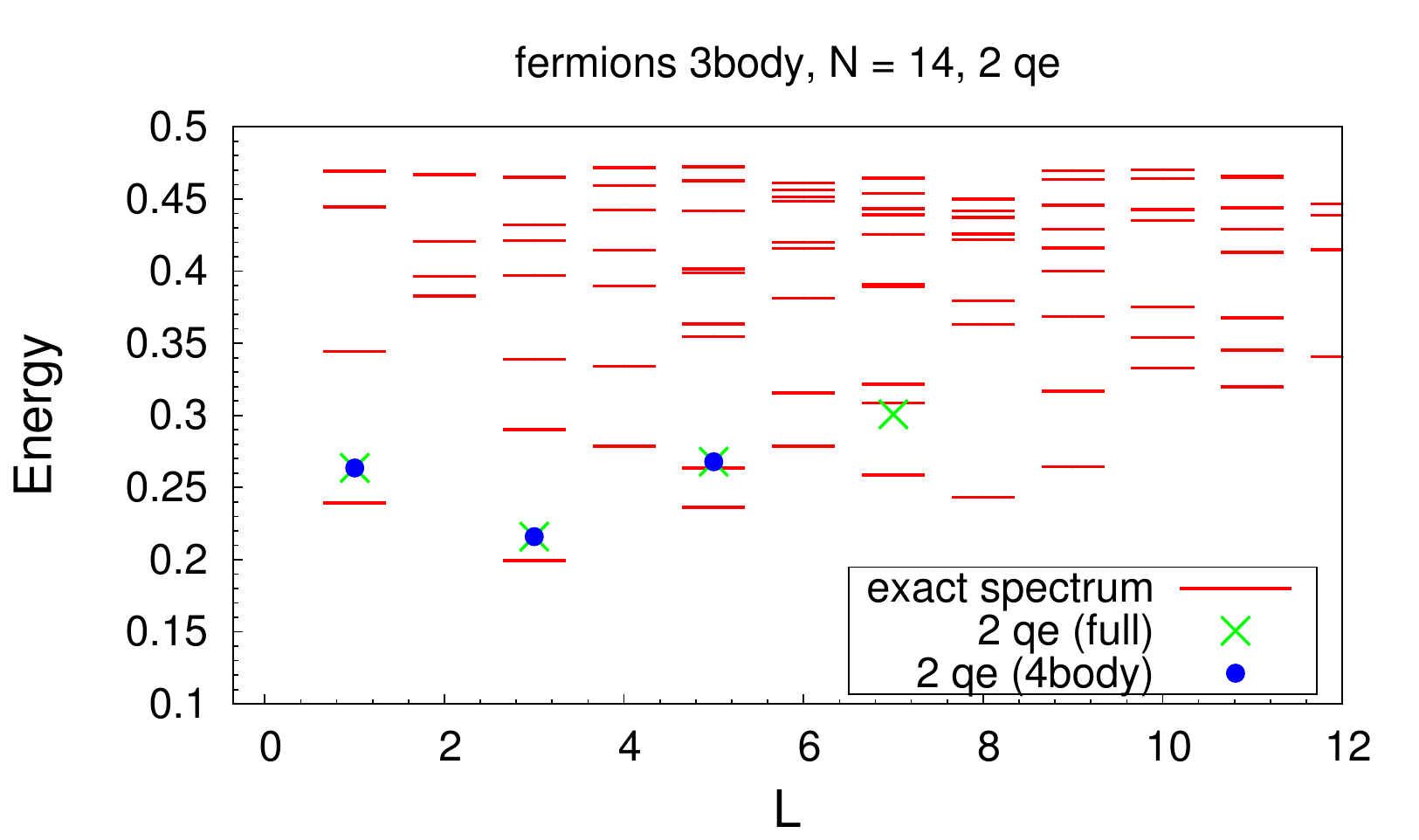}
\includegraphics[width= 8cm]{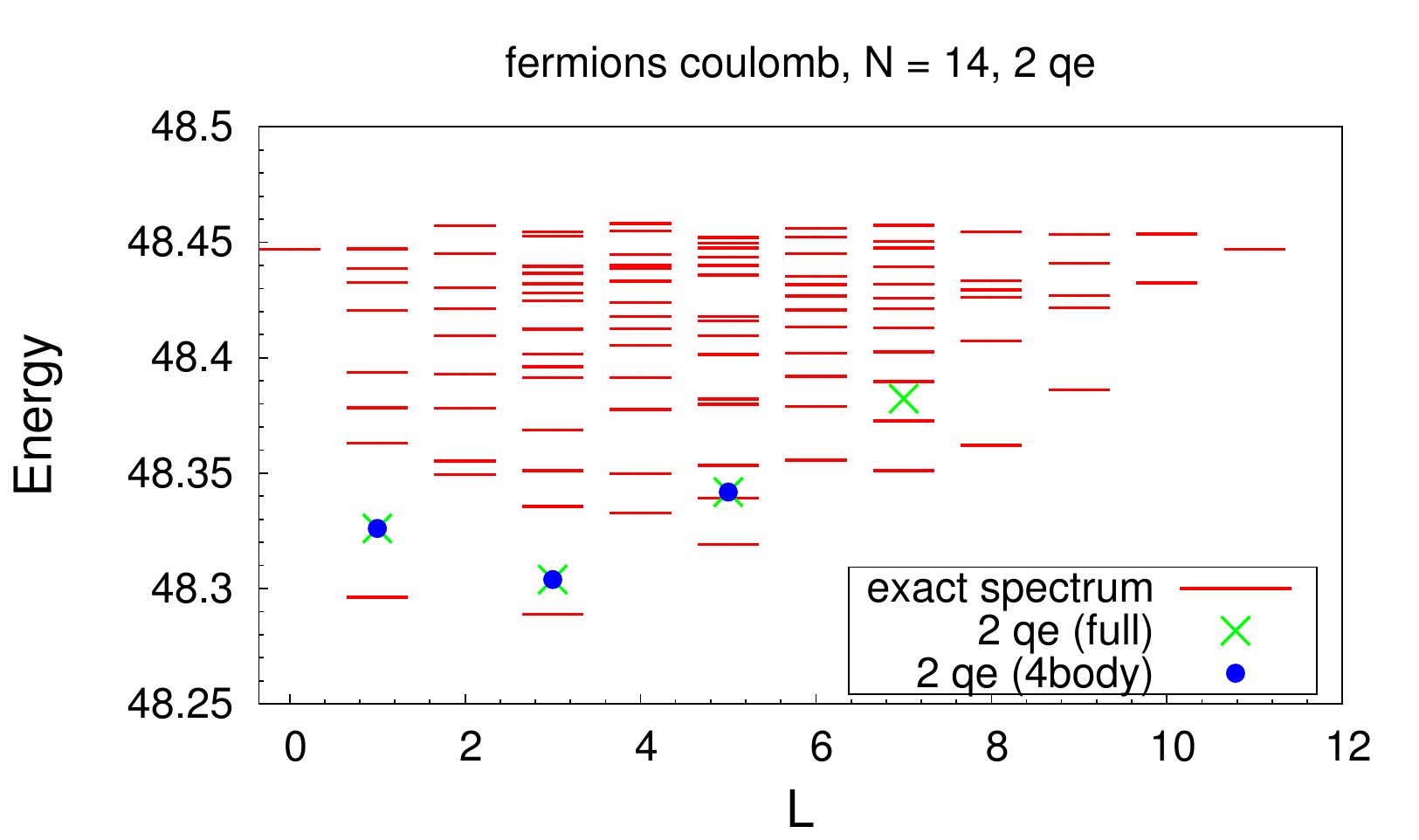}
\vspace*{-4mm}
\caption{
\textit{Top left}: Spectra for $N=16$ and $2$ quasielectrons ($N_{\phi} = 13$), of the Hamiltonian $\mathcal{H}_B^{(3)}$ in the full Hilbert space (dashes), in the full space of trial $2$-quasielectron states (crosses) and in the space of trial states that vanish when $4$ particles are brought to the same point (dots). 
\textit{Top right}: Spectra of $\mathcal{H}_B^{(2)}$ in the same spaces.\\
\textit{Bottom left}: Spectra of $\mathcal{H}_F^{(3)}$, for $N=14$ and $2$ quasielectrons ($N_{\phi} = 24$), in the analogous spaces for fermions. \textit{Bottom right}: Spectra of the second LL Coulomb Hamiltonian with $\delta V_1 =0.035$ in the same spaces.
}
\label{fig:spec_2qp}
\end{center}
\end{figure*}

In Fig.~\ref{fig:spec_2qp}, we show spectra for systems with $2$ quasielectrons.
The top panels of Fig.~\ref{fig:spec_2qp}
show the spectra of $\mathcal{H}_B^{(3)}$ and $\mathcal{H}_B^{(2)}$, in the full Hilbert space and in the space spanned by our trial states for $2$ quasielectrons, for $N=16$ bosons at $N_{\phi}=13$. The lower panels of Fig.~\ref{fig:spec_2qp} show the spectra of $\mathcal{H}_{F}^{(3)}$ and of the second LL Coulomb Hamiltonian with slightly modified $V_1$ pseudopotential, again in the full Hilbert space and in the space spanned by our trial states for $2$ quasielectrons, now for $N=14$ fermions at $N_{\phi}=23$. All panels also show the spectrum of the relevant Hamiltonians in the space of trial states for $2$ quasielectrons which have the additional property that they are zero modes of $\mathcal{H}_{B}^{(4)}$ (for bosons) or of $\mathcal{H}_{F}^{(4)}$ (for fermions). 

In all cases, the low-lying part of the spectrum in the space of trial wave functions is similar to the low lying part of the full spectrum. However, for bosons, the quasielectron states at the highest angular momentum (in this case $L = 8$) have anomalously large energy and do not obviously match anything in the exact spectra. For fermions the highest angular momentum states (now at $L=7$) also have relatively high energy, though the difference with the other trial states is not as pronounced as in the case of bosons. These high angular momentum states are also the only trial wave functions which do not vanish when $4$ particles are brought to the same point. More precisely, in the case of fermionic trial wave functions, the corresponding bosonic trial wave function does not vanish. Thus, it seems that the $4$-body vanishing property satisfied by the trial wave functions at lower $L$ may play a role in obtaining good agreement between trial states and low-energy states, as predicted by Haldane and Bernevig.\cite{Bernevig06}. Note that all trial wave functions vanish when $5$ particle positions coincide.  
We may also note that for bosons, the low-energy part of the spectrum of the model Hamiltonian $\mathcal{H}_B^{(3)}$ is obviously better reproduced than that of the realistic  $\mathcal{H}_B^{(2)}$ Hamiltonian. For fermions, the quality of approximation of the realistic and model Hamiltonians is comparable.

\begin{figure*}[htb]
\begin{center}
\vspace*{-1mm}
\includegraphics[width= 8cm]{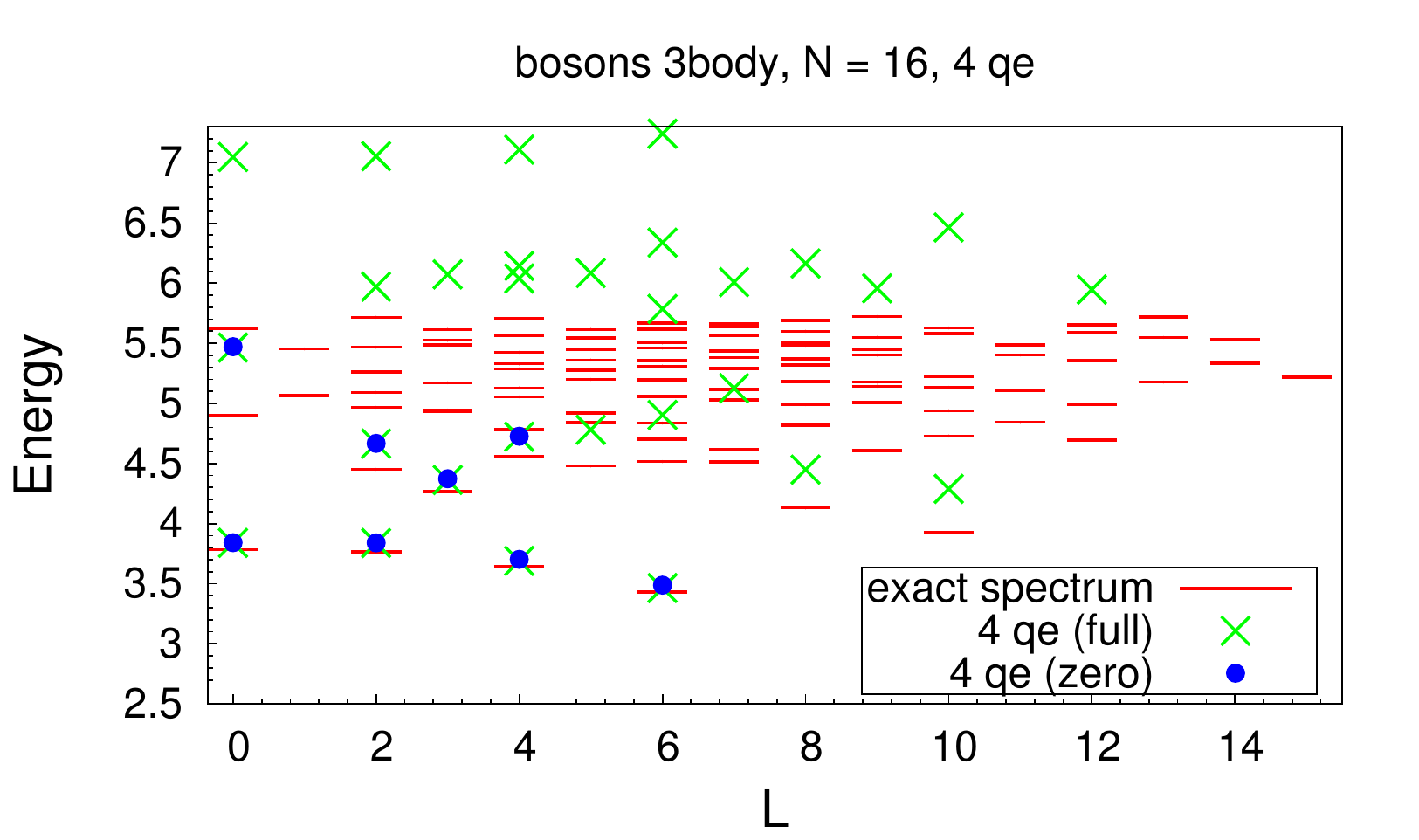}
\includegraphics[width= 8cm]{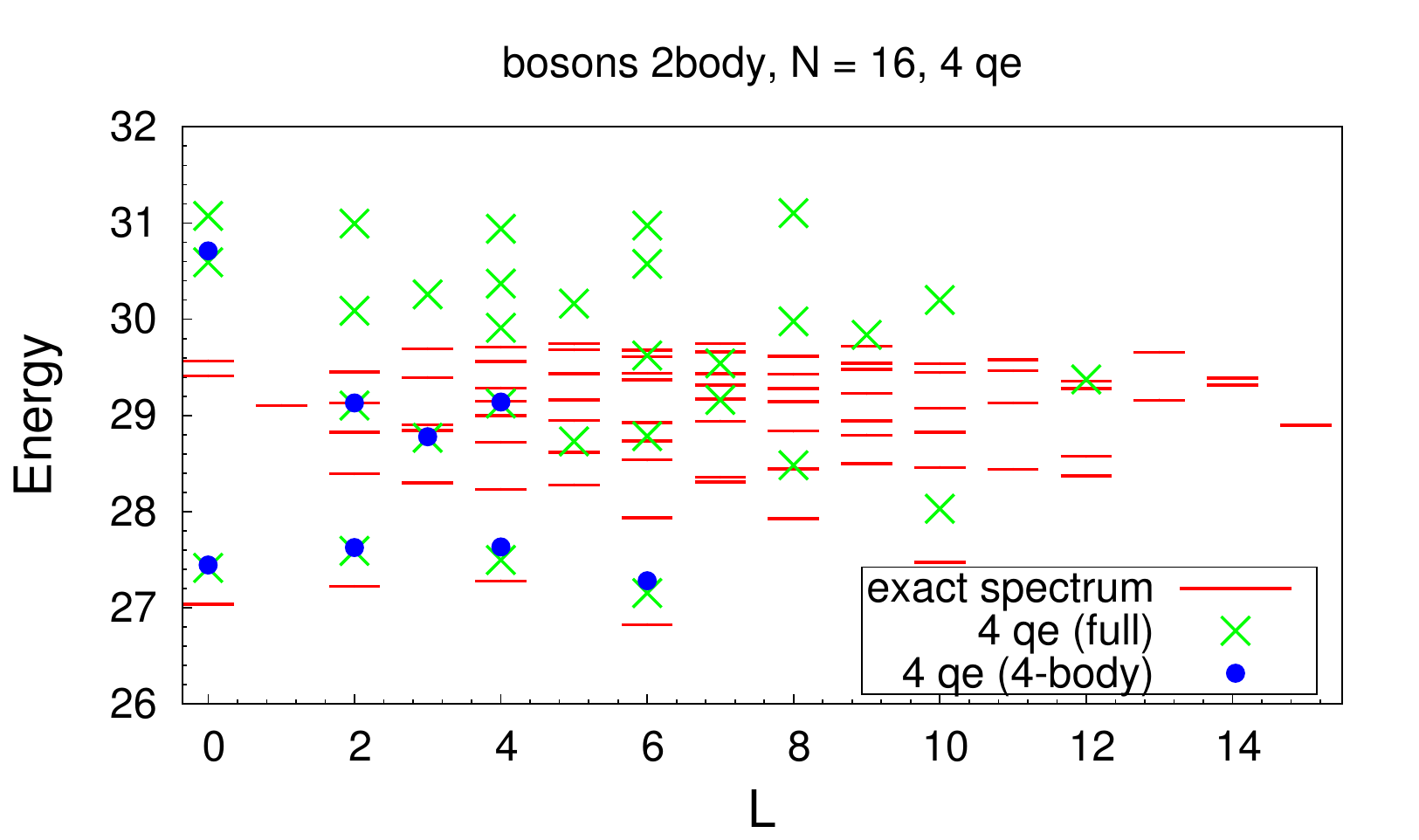}
\includegraphics[width= 8cm]{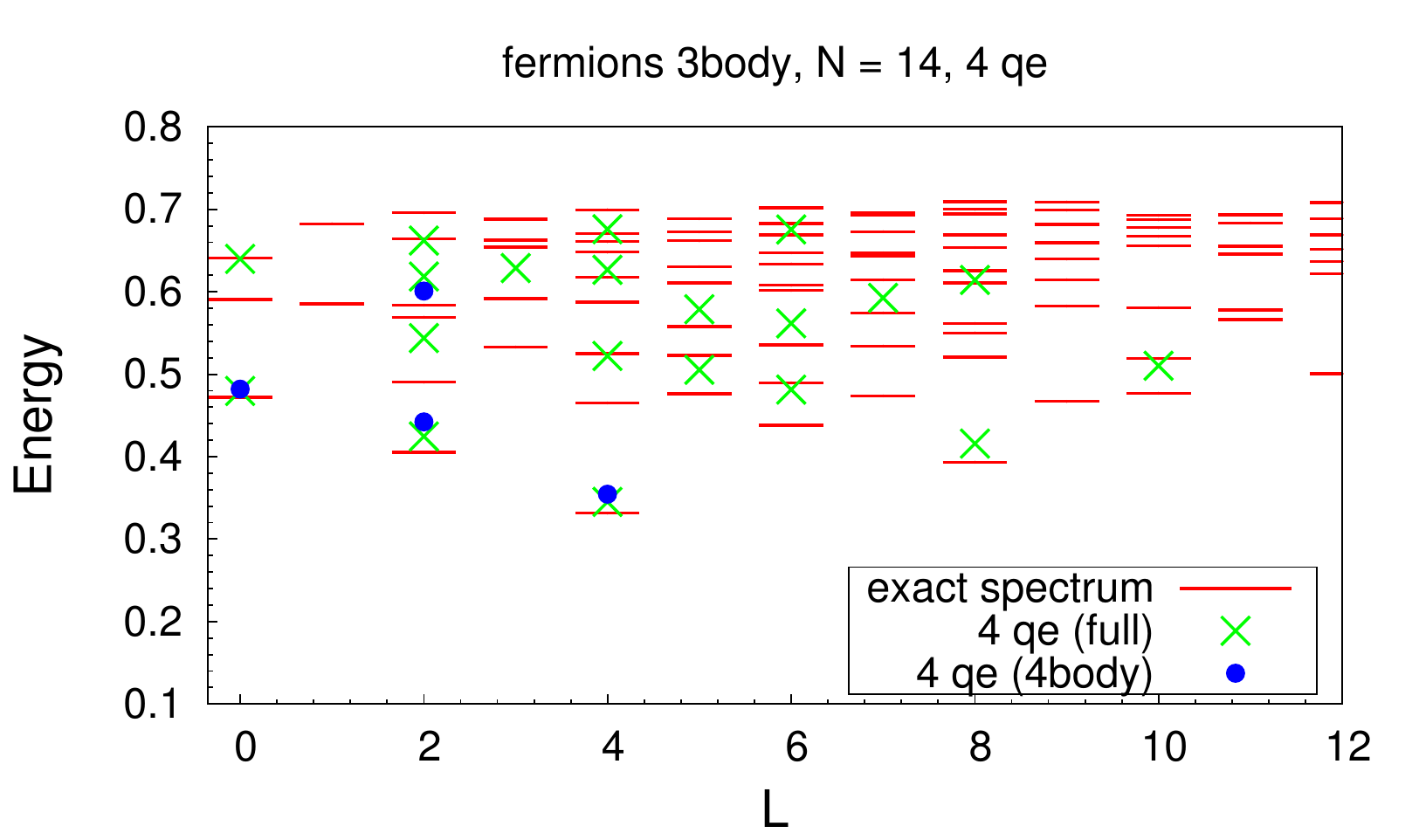}
\includegraphics[width= 8cm]{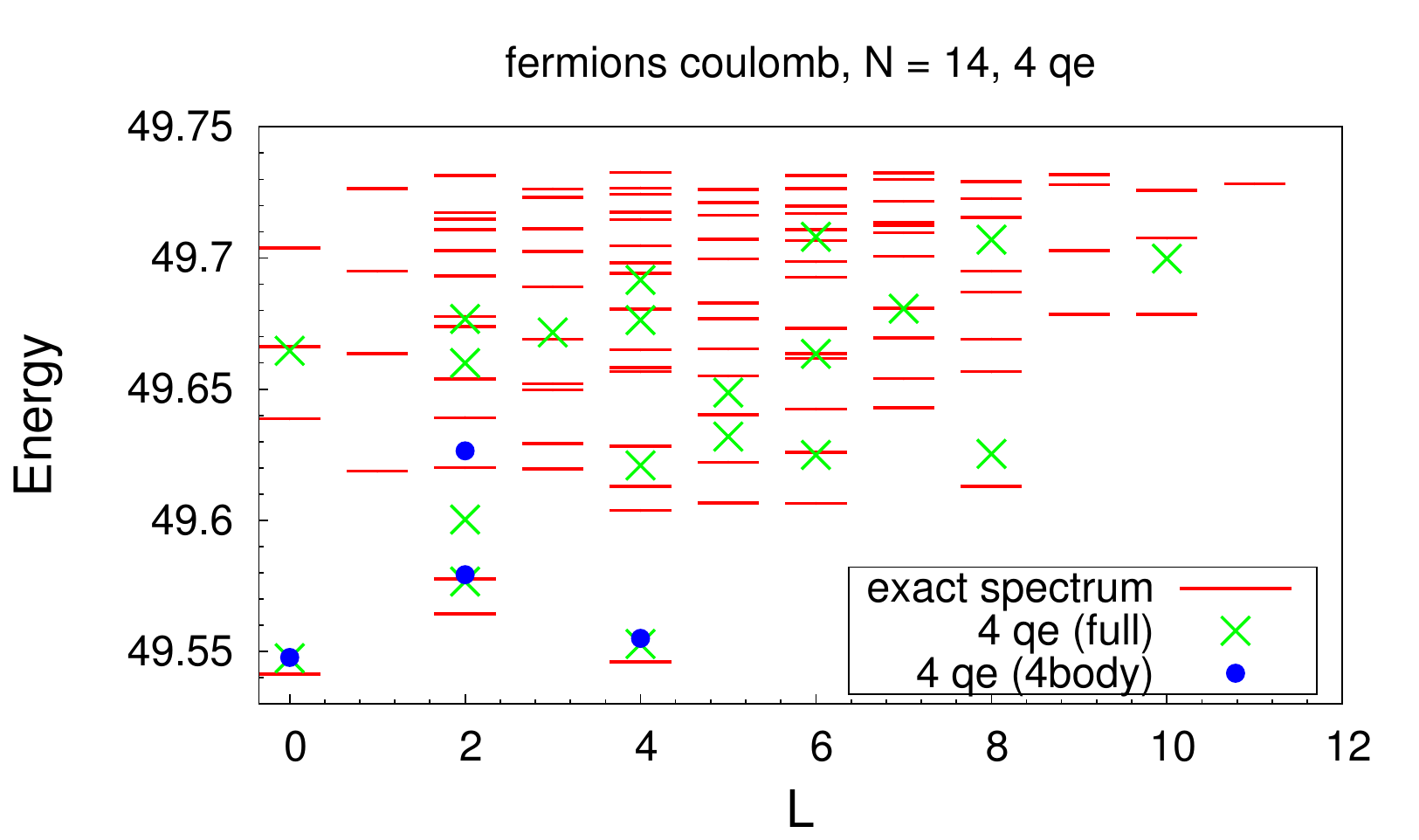}
\vspace*{-4mm}
\caption{
\textit{Top left}: Spectra for $N=16$ and $4$ quasielectrons ($N_{\phi} = 12$), of the Hamiltonian $\mathcal{H}_B^{(3)}$ in the full Hilbert space, in the space of trial $4$-quasielectron states and in the space of trial states that vanish when $4$ particles positions coincide. 
\textit{Top right}: Spectra of $\mathcal{H}_B^{(2)}$ in the same spaces.
\textit{Bottom left}: Spectra of $\mathcal{H}_F^{(3)}$, for $N=14$ and $4$ quasielectrons ($N_{\phi} = 23$), in the analogous spaces for fermions. 
\textit{Bottom right}: Spectra of the second LL Coulomb Hamiltonian with $\delta V_1 =0.035$ in the same spaces.
}
\label{fig:spec_4qp}
\end{center}
\end{figure*}

We have also calculated overlaps between the trial states and the corresponding low lying states of the Hamiltonian. The results for bosonic states can be found in Table~\ref{overlap_boson} and the results for fermions in Table~\ref{overlap_fermion}. In both cases, the results for $2$-quasielectron states are in the first two columns. When we compare our bosonic trial wave functions for $2$ quasielectrons to the lowest $\mathcal{H}_B^{(3)}$ eigenstates, the overlaps are all over $0.93$, except for the $L = 8$ state for which the overlap is $0.3$. The overlaps with the lowest $\mathcal{H}_B^{(2)}$ eigenstates are lower across the board, except at the highest angular momentum, but there the overlap is still low at $0.4$. This result is consistent with the fact that the spectrum of $\mathcal{H}_B^{(2)}$ is reproduced quite a bit worse than that of $\mathcal{H}_B^{(3)}$ (see figure ~\ref{fig:spec_2qp}). For fermions, the overlaps of the trial wave functions with the spectrum of the model Hamiltonian $\mathcal{H}_F^{(3)}$ are not as high as the overlaps for bosons, with the lowest overlap equal to $0.83$ if we exclude the $L=7$ state which is not a zero mode of $\mathcal{H}_F^{(4)}$. However, in the fermionic case, the overlaps for the realistic Coulomb Hamiltonian are comparable to those for the model Hamiltonian. Also, the potentially anomalous state at the highest $L$-value (here $L=7$) has considerably better overlap in the case of fermions, reaching $0.62$ for the Coulomb Hamiltonian. Nevertheless it still has much lower overlap than the trial states at lower $L$-values.
\begin{widetext}
\begin{center}
\begin{table}[t]
\begin{center}
\begin{tabular}{|c|c|c|c|c|c|c|c|c|}
\hline
$L$ & $\mathcal{H}_B^{(3)}$, $2qe$ & $\mathcal{H}_B^{(2)}$, $2qe$
&$\mathcal{H}_B^{(3)}$, $4qe$ &$\mathcal{H}_B^{(2)}$, $4qe$
&$\mathcal{H}_B^{(3)}$, $4qe$ $4$body&$\mathcal{H}_B^{(2)}$, $4qe$
$4$body &$\mathcal{H}_B^{(3)}$, $1 ex$ & $\mathcal{H}_B^{(2)}$, $1 ex$
\\\hline
0&0.985&0.915&0.619&0.527&0.754&0.613&-&- \\
1&-&-&-&-&-&-&-&-\\
2&0.970& 0.836&0.576&0.521&0.895&0.770&0.273&0.181\\
3&-&-&0.580&0.469&0.927&0.772&0.917&0.541\\
4&0.969&0.734&0.542&0.490&0.891&0.546&0.972&0.480\\
5&-&-&0.510&0.497&-&-&0.984&0.798\\
6&0.935&0.672&0.520&0.478&0.964&0.809&0.985&0.760\\
7&-&-&0.378&0.348&-&-&0.984&0.798\\
8&0.312&0.410&0.412&0.397&- &-&0.983&0.607\\
9&-&-&0.415&0.348&-&-&-&-\\
10&-&-&0.470&0.463&-&-&-&-\\
11&-&-&-&-&-&-&-&-\\
12&-&-&0.013&0.167&-&-&-&-\\\hline
\end{tabular}
\caption{\label{overlap_boson}Overlap between the space spanned by
quasielectron states for $N=16$ and the corresponding low energy space with respect
to the $\mathcal{H}_B^{(3)}$ and $\mathcal{H}_B^{(2)}$ Hamiltonian in
each $L$ sector. 
'$4$body' indicates that only states are taken into account that vanish when $4$ particles are at the same position. 
For $2$ quasielectron states, the only state that does not have this vanishing properties is the state at $L = 8$. 
We can notice that overlaps with this state are much smaller than the other ones. A dash
means that there is no state for the corresponding $L$ value.}
\end{center}
\end{table}
\begin{table}[bht]
\begin{center}
\begin{tabular}{|c|c|c|c|c|c|c|c|c|}
\hline
$L$ &$\mathcal{H}_F^{(3)}$, $2qe$& $ \widetilde{\mathcal{H}_{C}}$, $2qe$& $\mathcal{H}_F^{(3)}$, $4qe$  &$\widetilde{\mathcal{H}_{C}}$, $4qe$ &$\mathcal{H}_F^{(3)}$, $4qe$ $4$body &$\widetilde{\mathcal{H}_{C}}$, $4qe$ $4$body &$\mathcal{H}_F^{(3)}$, $1 ex$ &$\widetilde{\mathcal{H}_{C}}$, $1 ex$ \\ \hline
0&-&-&0.836&0.839&0.955&0.973&-&-\\
1&0.903&0.792&-&-&-&-&-&-\\
2&-&-&0.791&0.757&0.629&0.732&0.133&0.026\\
3&0.929&0.910&0.571&0.003&-&-&0.471&0.214\\
4&-&-&0.638&0.568&0.911&0.947&0.883&0.710\\
5&0.830&0.840&0.679&0.690&-&-&0.958&0.821\\
6&-&-&0.617&0.597&-&-&0.947&0.738\\
7&0.593&0.621&0.299&0.569&-&-&0.943&0.817\\
8&-&-&0.660&0.473&-&-&-&-\\
9&-&-&-&-&-&-&-&-\\
10&-&-&0.835&0.708&-&-&-&-\\
\hline
\end{tabular}
\caption{\label{overlap_fermion}Overlap between the space spanned by fermionic quasielectron states for $N=14$ and the corresponding space of low energy states with respect to $\mathcal{H}_F^{(3)}$ and $\widetilde{\mathcal{H}_{C}}$, the Coulomb interaction in the second LL Hamiltonian $+\delta V_1 = 0.035$ in each $L$ sector. A dash means that there is no state for the corresponding $L$ value.}
\end{center}
\end{table}
\end{center}
\end{widetext}
We similarly investigated systems with $4$ quasielectrons. 
In Fig.~\ref{fig:spec_4qp}, we show the spectra for these systems.
The top panels again show the spectra of $\mathcal{H}_B^{(3)}$ and $\mathcal{H}_B^{(2)}$, in the full Hilbert space and in the space spanned by our trial states for $4$ quasielectrons, for $N=16$ bosons at $N_{\phi}=12$. The lower panels show the spectra of $\mathcal{H}_{F}^{(3)}$ and of the second LL Coulomb Hamiltonian for $N=14$ fermions at $N_{\phi}=23$. In both cases, the number of trial states generated is much higher than for $2$ quasielectrons. Only a few of these states vanish when a cluster of $4$ particles is formed. As in the case of $2$ quasielectrons, these states belong to the low energy part of the spectra. Overlaps between fixed $L$ subspaces spanned by $4$-quasielectron trial states and subspaces of lowest energy states of the Hamiltonians at the same $L$-value are given in the third and fourth columns of Tables~\ref{overlap_boson} and \ref{overlap_fermion}.  Overlaps between the subspaces of $4$-body vanishing states and subspaces of lowest energy states of the Hamiltonians are shown in the fifth and sixth columns of these tables. The overlaps for the $4$-body vanishing states are considerably higher than those for the full set of trial states and indeed it is also easy to see from the spectra that the $4$-body vanishing states are among the lowest energy trial states and certainly give a very economical description of the lowest-energy part of the full spectrum. For bosons there even appears to be a low energy `band' of states at $L=0$, $L=2$, $L=4$ and $L=6$ which can be described by a subset of the trial wave functions with the $4$-body vanishing property. On the other hand it is obvious from the spectra, especially from those for fermions, that the full set of trial wave functions does give a reasonable description of a much larger part of the low energy spectrum than the $4$-body vanishing states. 

\begin{figure*}[htb]
\begin{center}
\vspace*{-3mm}
\includegraphics[width= 8cm]{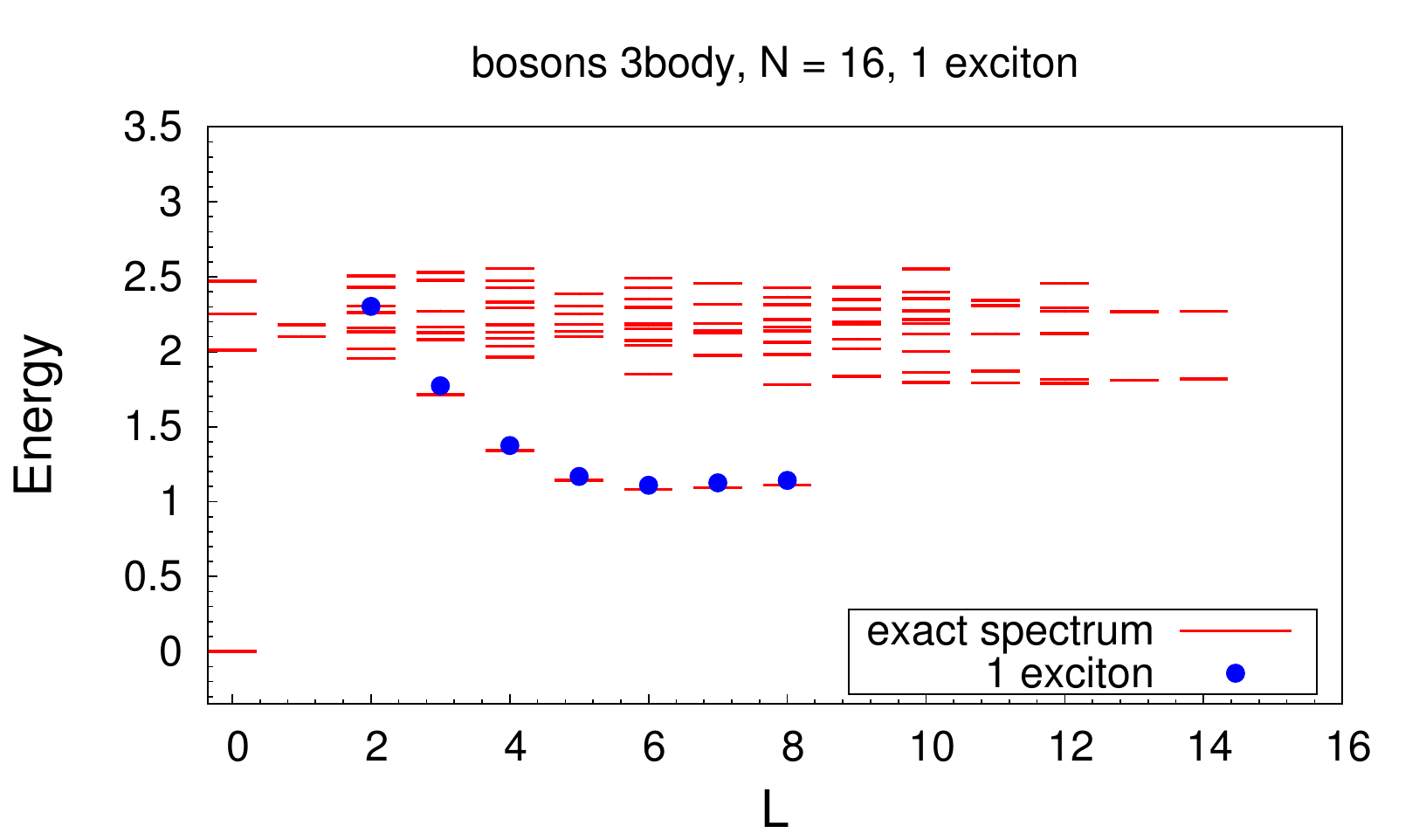}
\includegraphics[width= 8cm]{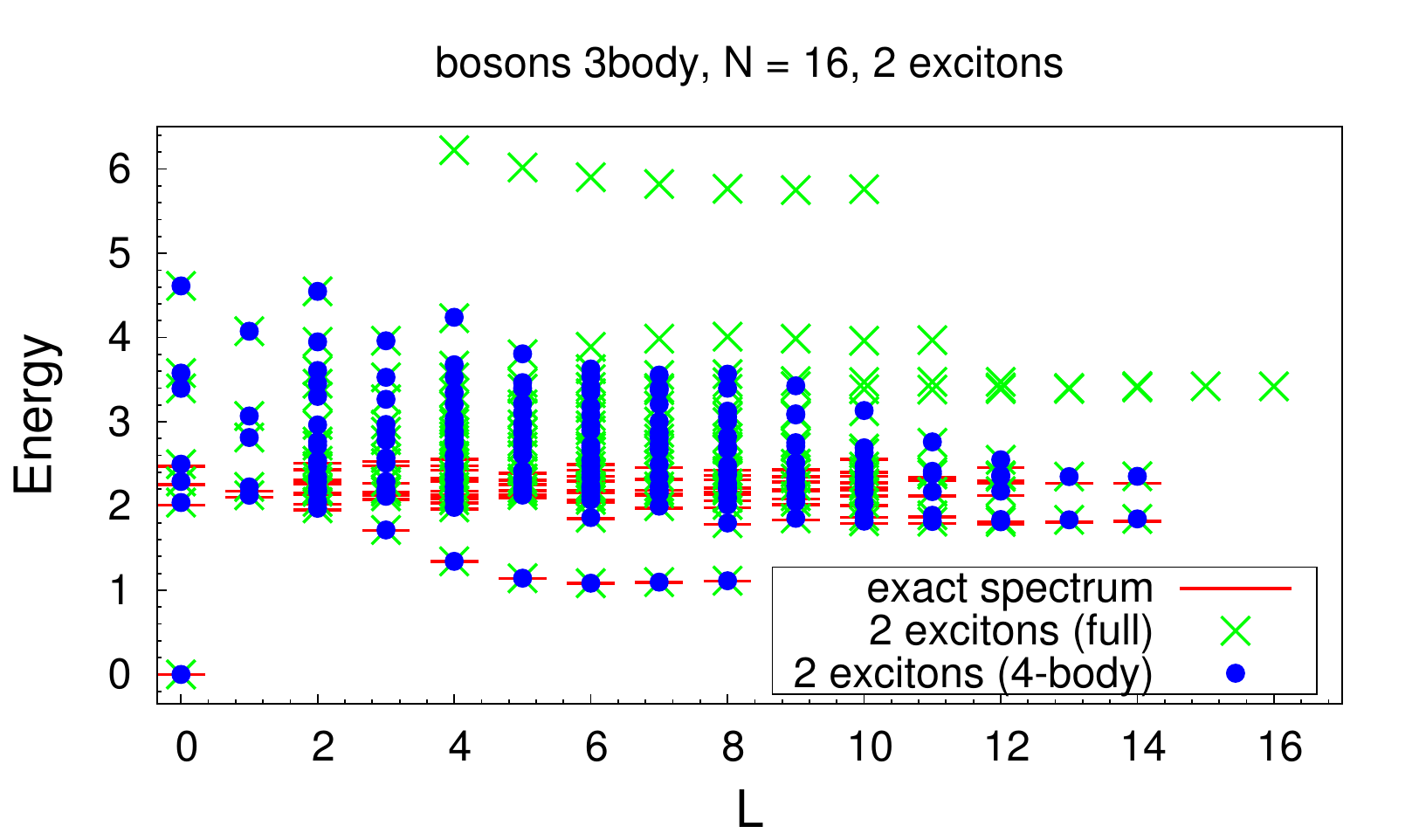}
\includegraphics[width= 8cm]{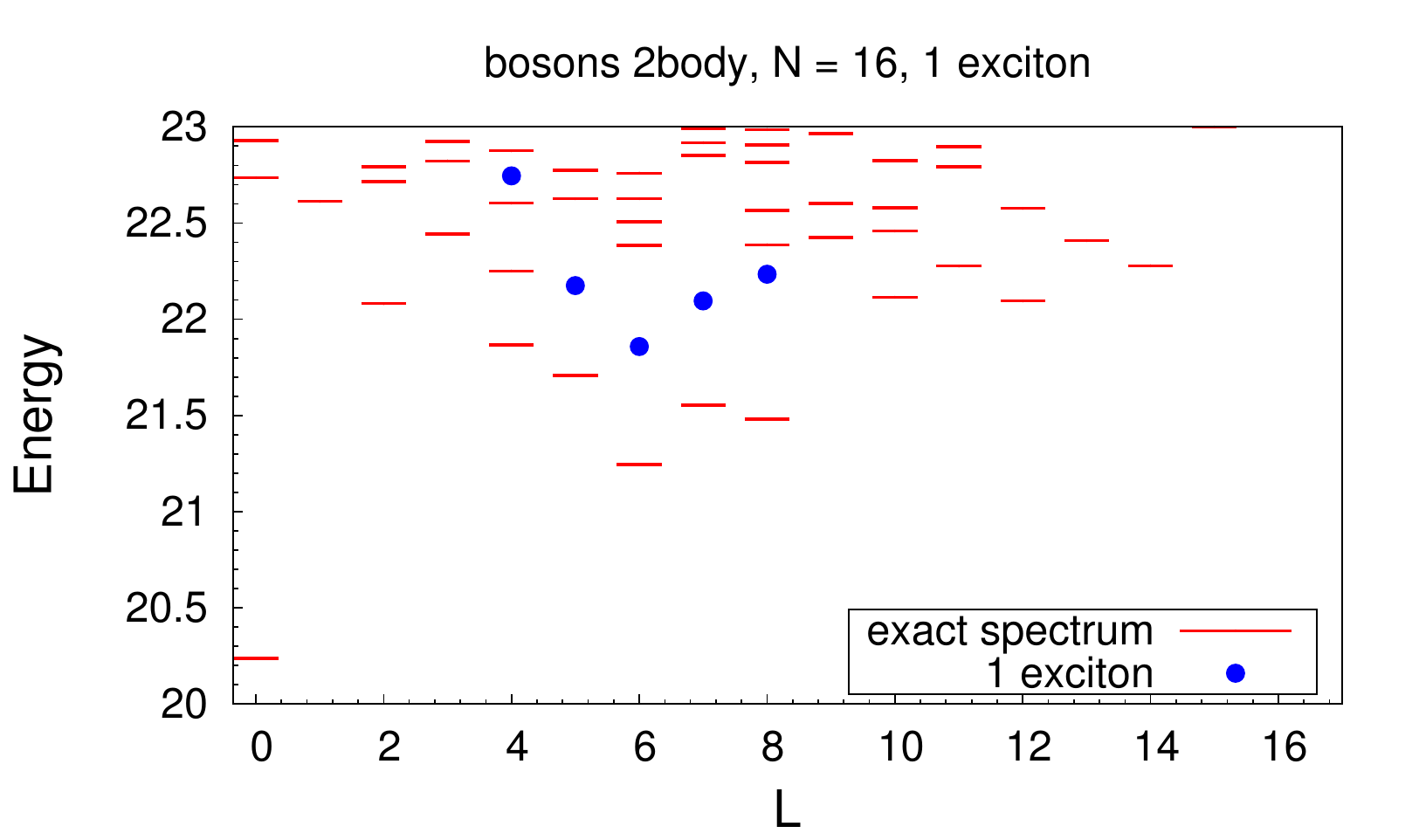}
\includegraphics[width= 8cm]{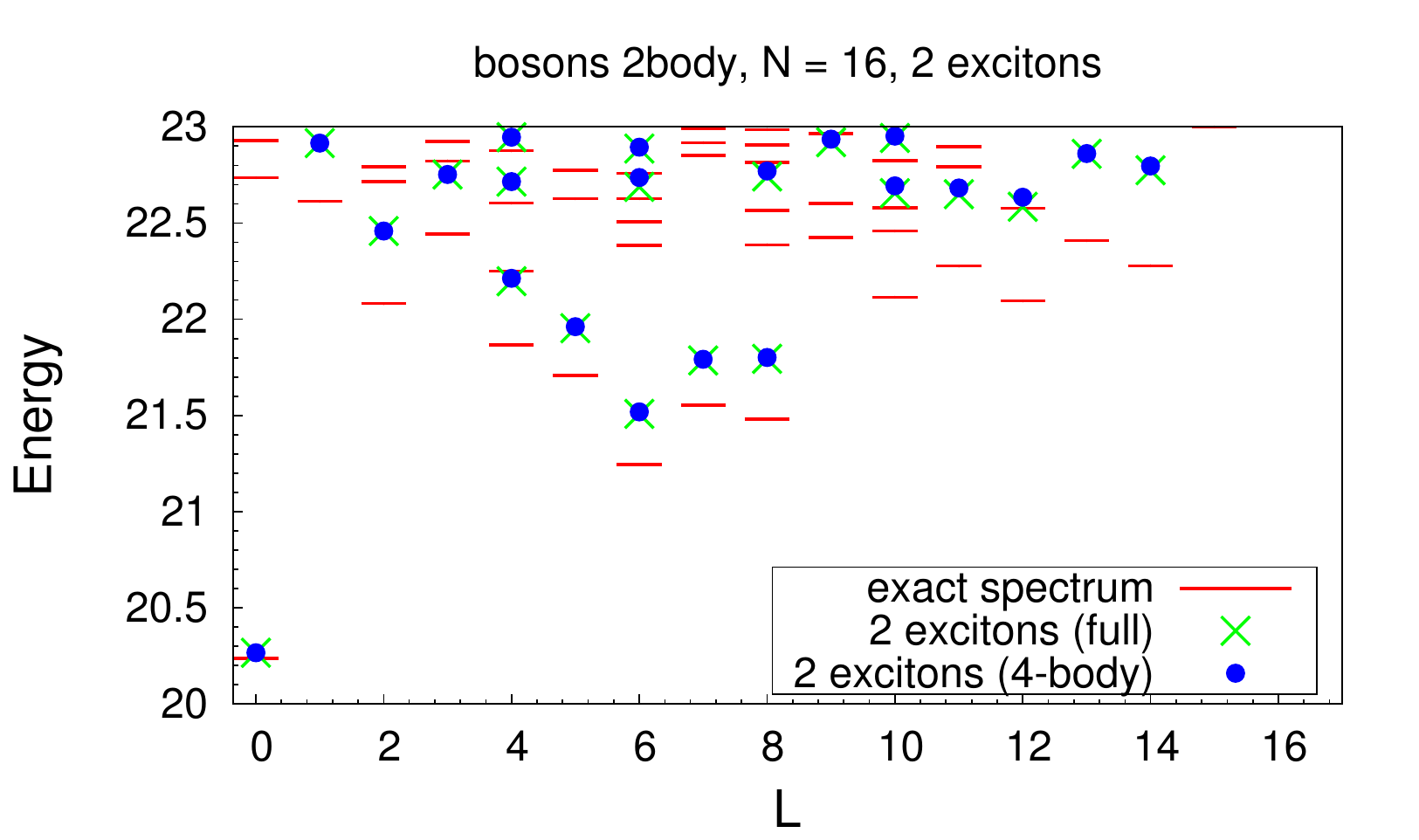}
\vspace*{-4mm}
\caption{
Spectra for bosons at $N = 16$ and $N_{\phi} = 14$. 
\textit{Top left}: 
Spectrum of $\mathcal{H}_B^{(3)}$ in the full Hilbert space (dashes) and in the space spanned by our one exciton trial states (dots).
\textit{Bottom left}: Spectra of $\mathcal{H}_B^{(2)}$ in the same spaces.
\textit{Top right}: Spectra of $\mathcal{H}_B^{(3)}$ in the full Hilbert space, in the space of our $2$-exciton trial states (crosses) and in the space of $2$-exciton trial wave functions which are zero modes of $\mathcal{H}_{B}^{(4)}$ (dots). 
\textit{Bottom right}: Spectra of $\mathcal{H}_B^{(2)}$ in the same spaces.
}
\label{fig:spec_exc_bosons}
\end{center}
\end{figure*}

\begin{figure*}[htb]
\begin{center}
\vspace*{-1mm}
\includegraphics[width= 8cm]{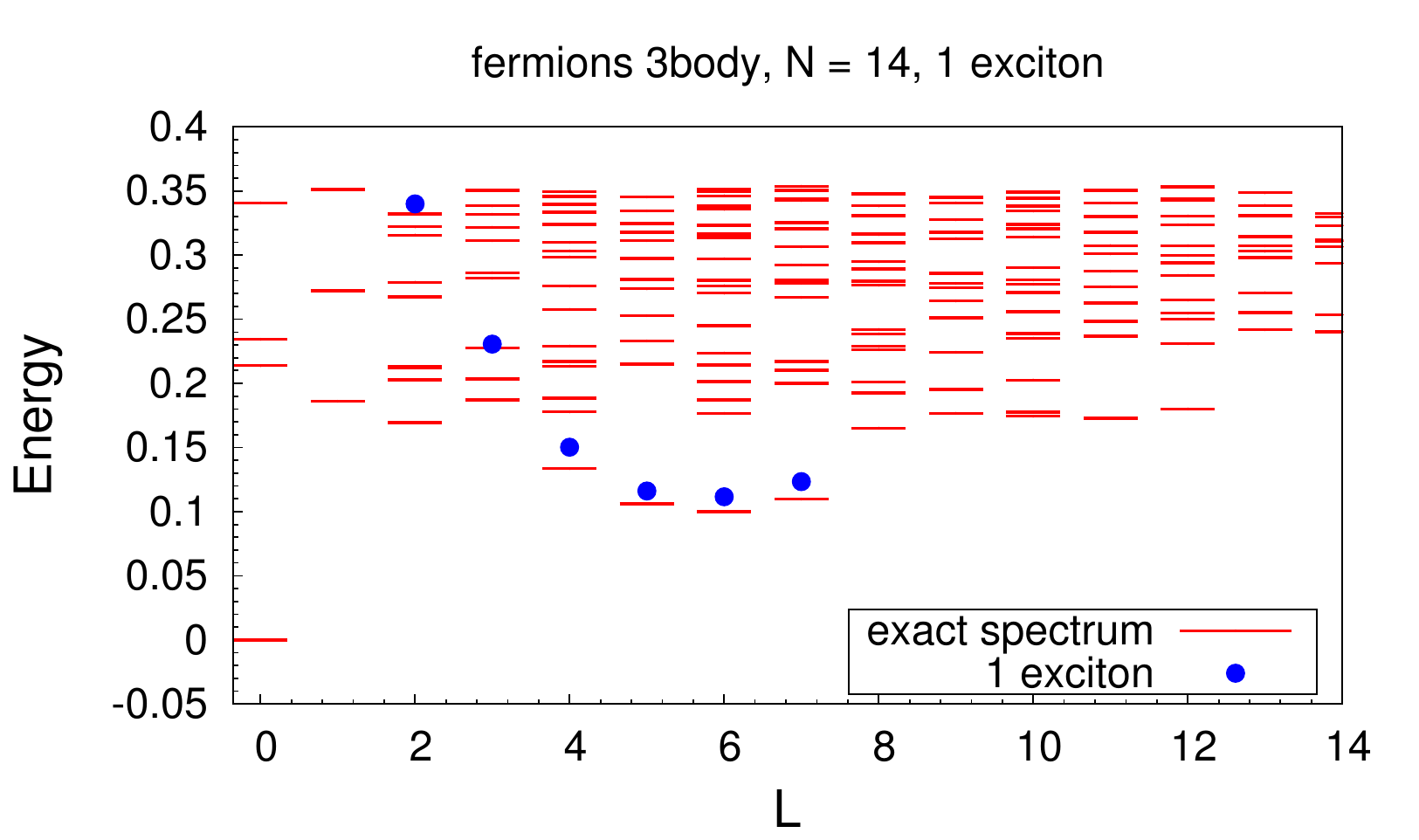}
\includegraphics[width= 8cm]{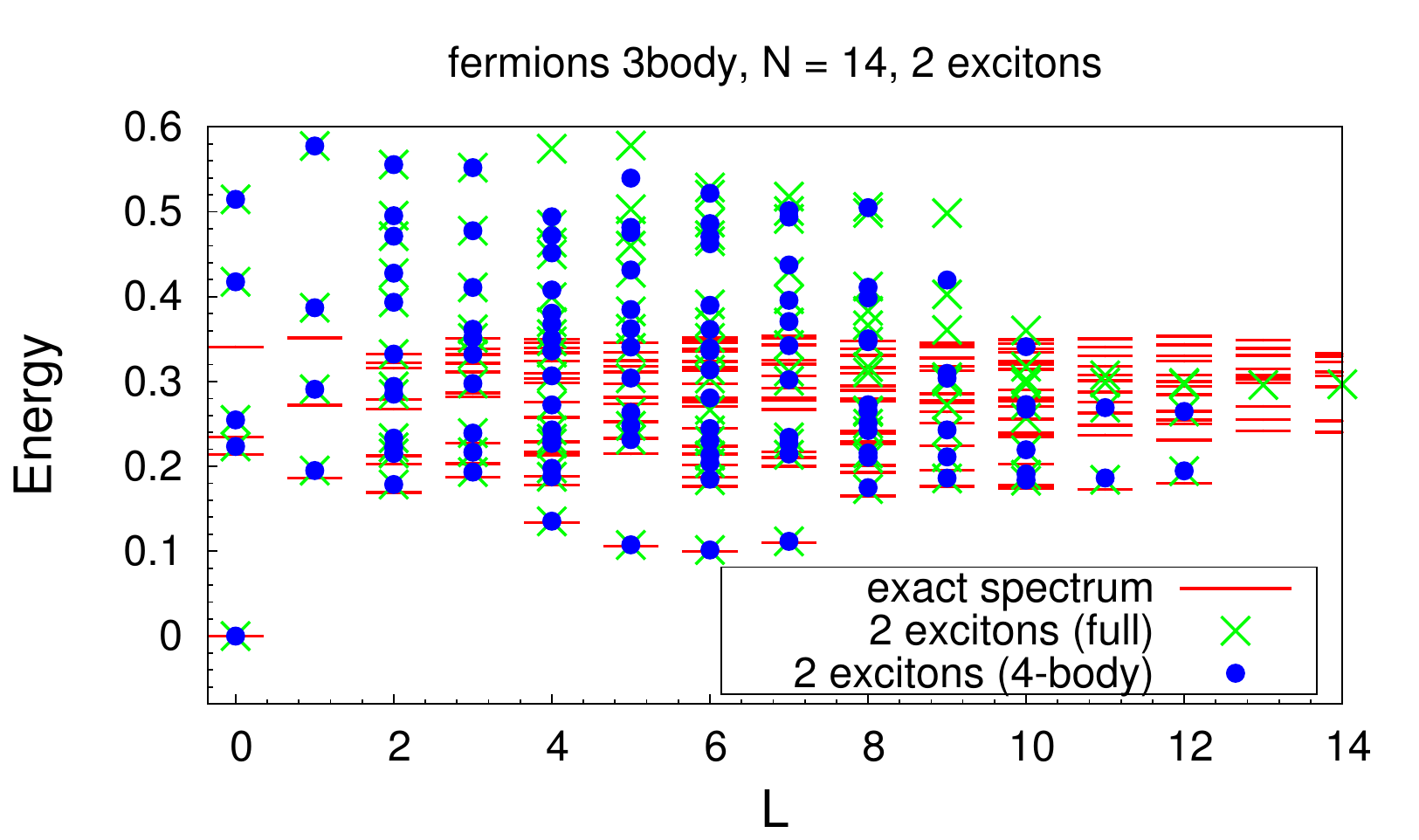}
\includegraphics[width= 8cm]{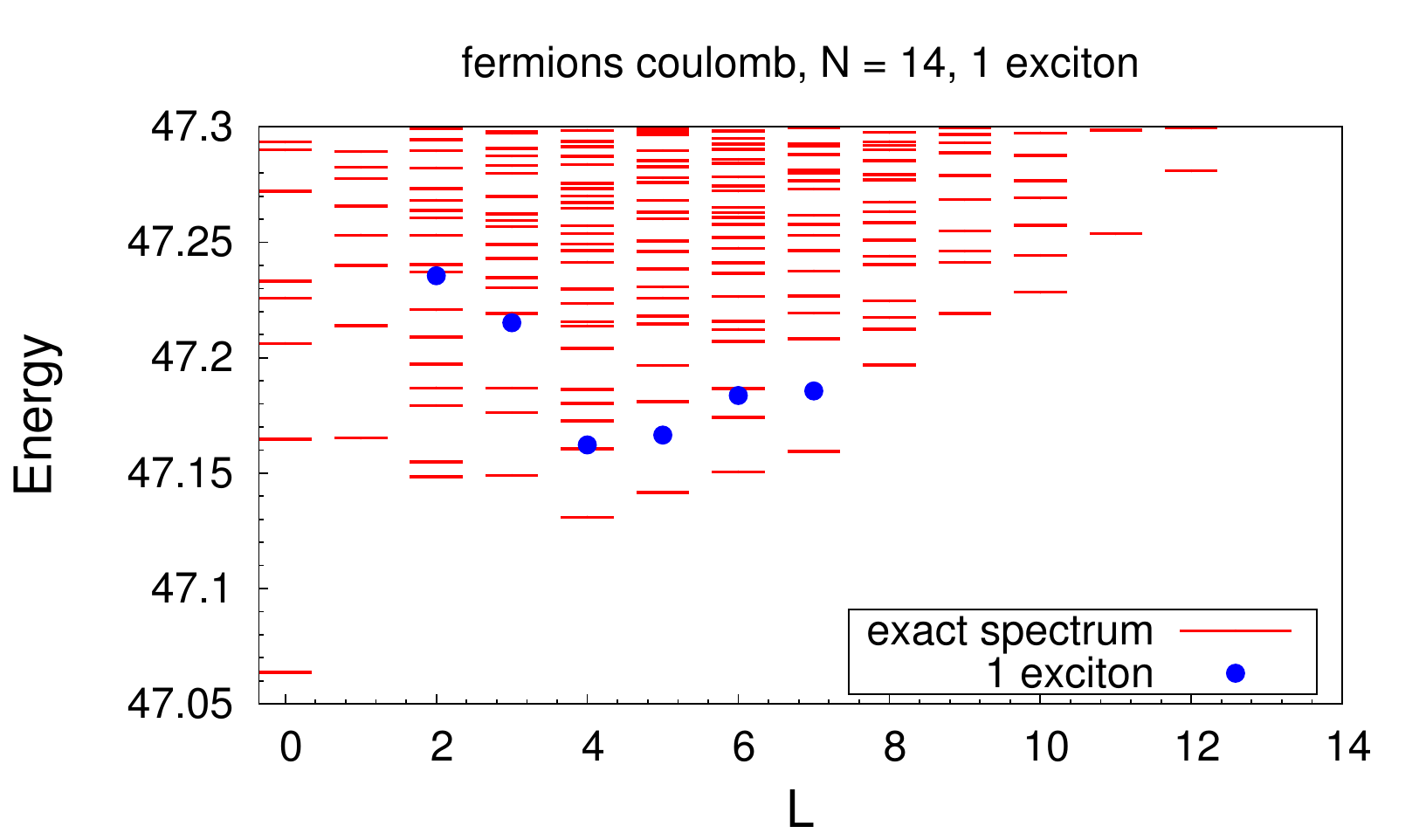}
\includegraphics[width= 8cm]{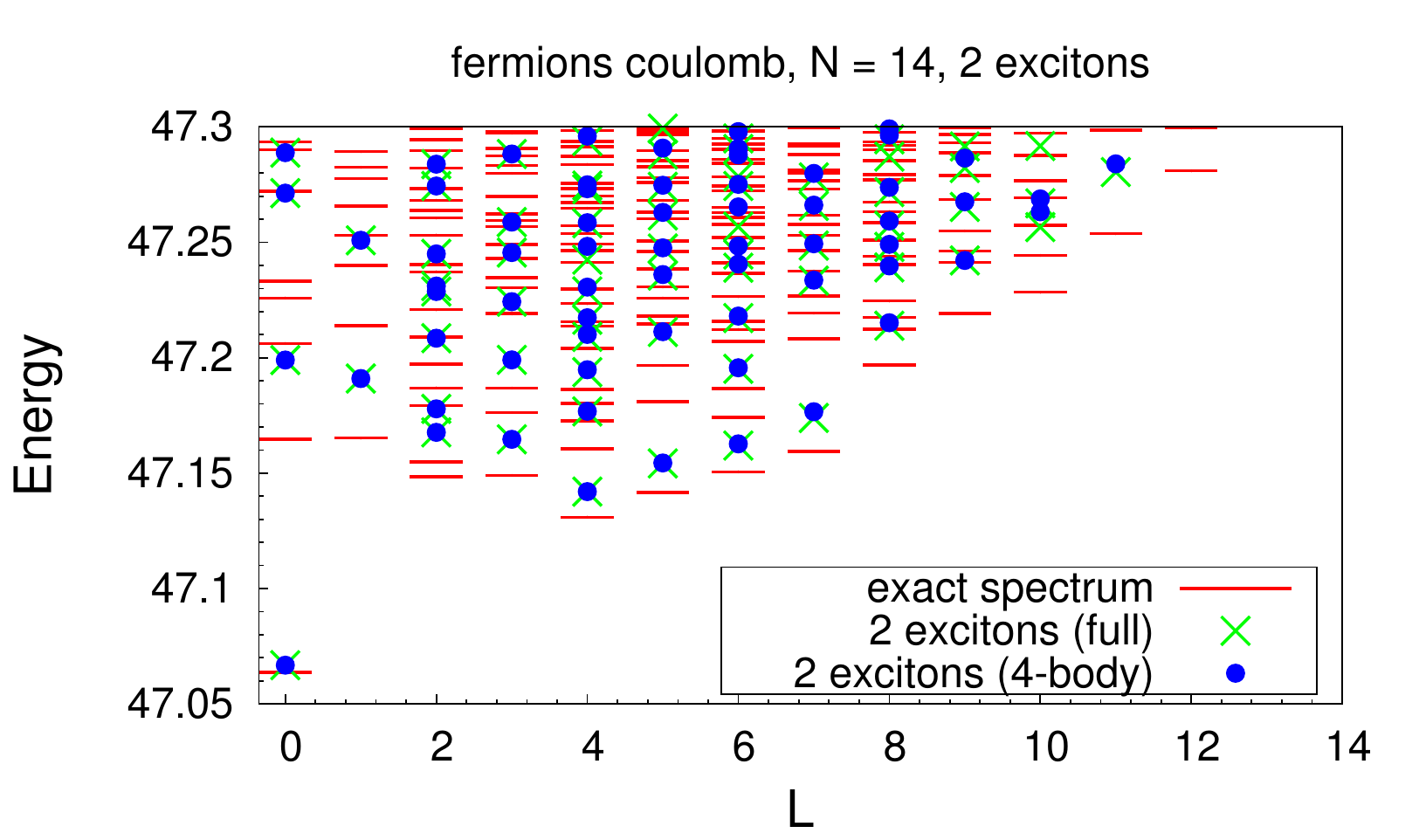}
\vspace*{-4mm}
\caption{
Spectra for fermions at $N = 14$ and $N_{\phi} = 25$. 
\textit{Top left}: Spectra of $\mathcal{H}_{F}^{(3)}$ in the full Hilbert space (dashes) and in the space spanned by our one exciton trial states (dots).
\textit{Bottom left}: Spectra of the second LL Coulomb interaction with $\delta V_1 =0.035$ in the same spaces.
\textit{Top right}: Spectra of $\mathcal{H}_{F}^{(3)}$ in the full Hilbert space, in the space of our $2$-exciton trial states (crosses) and in the space of $2$-exciton trial wave functions which are zero modes of $\mathcal{H}_{F}^{(4)}$ (dots). 
\textit{Bottom right}: Spectra of the second LL Coulomb interaction with $\delta V_1 =0.035$ in the same spaces.}
\label{fig:spec_exc_fermions}
\end{center}
\end{figure*}

We have examined our construction of exciton states by the same means. We consider the two lowest effective cyclotron energies here, i.e.~exciton states with effective composite fermion energy up to $2$ in units of the effective cyclotron energy of the composite fermions.  The spectra of the different Hamiltonians in the spaces spanned by the $1$-exciton and $2$-exciton trial states and in the full Hilbert space are shown in Fig.~\ref{fig:spec_exc_bosons} for bosons and in Fig.~\ref{fig:spec_exc_fermions} for fermions. Overlaps between single exciton states and low energy states in the full Hilbert space are given in Table~\ref{overlap_boson} for bosons and in Table~\ref{overlap_fermion} for fermions. 

Single excitons (with effective energy $1$) are obtained when one of the liquids is in the Laughlin state while the second is in a state excited over the Laughlin state with one exciton. These states naturally vanish when $4$ particles are brought to the same point (see Section~\ref{sec:twolayervanishing}). In the case of the model Hamiltonians $\mathcal{H}_B^{(3)}$ and  $\mathcal{H}_F^{(3)}$, the magneto-roton-like branch is remarkably well reproduced by these $1$-exciton states, both in the spectra and in the overlaps. For the realistic Hamiltonians the performance of the $1$-exciton trial wave functions is less impressive, but still quite reasonable. 

As discussed in Section~\ref{sec:num_result_counting}, all $2$-exciton states can be generated by considering only the case when both liquids are in a $1$-exciton state. In this case, the a priori vanishing property is a $5$-body cancellation. However, most of the states also vanish when $4$ particle are brought to the same point (see Section~\ref{sec:num_result_counting} for details). The different spectra in the space of $2$-exciton states with the $4$-body vanishing property are also shown in Figs.~\ref{fig:spec_exc_bosons} and \ref{fig:spec_exc_fermions}. Due to the large number of trial states generated and the large number of exact low energy states of the various Hamiltonians involved, it is difficult to make a meaningful quantitative comparison of the different spectra, and so we have not listed overlaps for $2$-exciton states. However, we can see from the spectra that our construction behaves well with respect to the energy of model and realistic Hamiltonians, in the sense that, as we consider more trial states we manage to describe more of the lower energy part of the spectra.

\section{Discussion}
\label{sec:discussion}

We have proposed trial wave functions for charged and neutral excitations of the MR Pfaffian phase, tested these numerically and compared our construction to existing proposals. 

Despite what seems like a very different method of construction, we find that our wave functions  are precisely the angular momentum eigenstates which appear in the trial wave functions for localized quasiparticles and excitons proposed by Hansson et al.\cite{Hansson09}. On comparing with the trial states based on vanishing properties proposed by Haldane and Bernevig\cite{Bernevig06}, we find that our single exciton wave functions coincide with theirs, but our  quasielectron wave functions do not. Finally, our construction is very similar to a CF based construction by Sreejith et al.\cite{Sreejith11_PRL_107} and we show that the difference can be viewed as a change in the LLL projection.  

As one of our main results, we find that the counting of large $L$ multiplets of independent trial wave functions in our construction is the same for quasholes and quasielectrons, supporting the idea that quasiholes and quasielectrons are described by the same CFT and TQFT. This result is in disagreement with the claim of Sreejith et al.~in Ref.~\onlinecite{Sreejith11_PRL_107} that the counting of quasielectron states is different from that of quasiholes. While the trial wave functions of Ref.~\onlinecite{Sreejith11_PRL_107} differ from ours in the details of the LLL-projection, this is not the explanation of the disagreement, as we have checked for small systems that, with the method detailed in Section~\ref{sec:trad_meth_impl}, we find the same quasielectron multiplet countings for both types of trial states. 

We also tested how well our trial wave functions reproduce the spectra of idealized $3$-body Hamiltonians and of more realistic Hamiltonians, both for bosons and for fermions.
We find that the low lying parts of the energy spectra for the $3$-body Hamiltonians are reproduced very well when diagonalizing these Hamiltonians in the spaces of trial states for quasielectrons and excitons. This is also reflected in high overlaps between the low lying eigenstates in the spectra, especially for $2$-quasielectron states and single exciton states. For bosons, the trial wave functions reproduce the spectra quite a bit less well in the case of the more realistic $\mathcal{H}_{B}^{(2)}$ Hamiltonian, though the agreement between the spectra in the full space and trial spaces and the overlaps are still reasonable. For fermions, it appears that the agreement between our trial wave functions and the eigenstates of the second Landau level Coulomb Hamiltonian (with slightly shifted $\delta V_1$ to stabilize a MR Pfaffian phase) is roughly equally good as the agreement between our trial states and the eigenstates of the model $3$-body Hamiltonian. 

It would be of interest to perform a detailed comparison between our results for the spectra and overlaps of the fermionic wave functions and those in Ref.~\onlinecite{Sreejith11_PRL_107}, but this is not completely straightforward. Sreejith et al.~use the Coulomb Hamiltonian without the shift in the pseudopotential $V_1$ which we use to stabilize the Pfaffian phase. Also, the overlaps presented in Ref.~\onlinecite{Sreejith11_PRL_107} are overlaps between the lowest energy state of the Hamiltonian in the space of trial wave functions at a given value of $L$ and the lowest energy state of the Hamiltonian in the full Hilbert space at the same value of $L$. We present instead the overlap (\ref{overlap}) between the full space of trial wave functions at a given $L$-value and a space of low energy eigenstates of the Hamiltonian which has the same dimension.  We think that our approach is more consistent to probe the accuracy of an excitation manifold construction. Nevertheless, we checked, up to the definition of the overlap (we use $|\braket{\Psi^1}{\Psi^2}|^2$ instead of  $|\braket{\Psi^1}{\Psi^2}|$), that our results are in agreement with those of Ref. \onlinecite{Sreejith11_PRL_107} for the $2$ and $4$ quasielectrons case with the $\mathcal{H}_{F}^{(3)}$ Hamiltonian.

While our trial wave functions do reproduce the low lying parts of the spectra with at least reasonable success for each of the Hamiltonians considered, it is also clear that many of the trial states have high variational energies and could be dispensed with. In fact, we observe that there is a subspace of our space of trial wave functions which typically have much better variational energies and overlaps compared to the other trial wave functions, especially in the case of bosons. This is the space of trial wave functions with $4$-body vanishing properties, or more precisely, the space of zero modes of $\mathcal{H}_{B}^{(4)}$ for bosons and the space of zero modes of $\mathcal{H}_{F}^{(4)}$ for fermions. This observation suggests that it is a good idea to construct trial wave functions 
based on their vanishing properties, a technique already employed by Bernevig and Haldane in Ref.~\onlinecite{Bernevig06}. It is important to note however, that one must employ more complicated vanishing requirements than those given by Bernevig and Haldane if one wishes to describe arbitrarily large numbers of quasielectrons. One reason for this is that, as one decreases the number of flux quanta $N_{\phi}$ in the system (at fixed $N$), from the MR ground state's flux of $N_{\phi}=N-2$ (for bosons), one eventually reaches values of $N_{\phi}$ where no states exist which vanish when $4$ particle positions coincide. This happens for bosons when $N_{\phi}<\frac{2}{3}N-2$, the flux of the bosonic $k=3$ Read-Rezayi state. 
One might argue that, in a real system, $N$ is very large and one would never have to worry about systems with this many quasielectrons.

However, in the small systems that are used in numerical studies, the number of $4$-body vanishing states can start decreasing already at a relatively modest number of quasielectrons (especially if further vanishing requirements are imposed) and thus one cannot hope to extract the correct large $N$ limit of the counting of multiplets of quasielectrons from numerical calculations on such states. Also, it must be noted that for fermions the difference in overlaps and energies between states with the $4$-body vanishing property and other low-lying trial states is less pronounced than for bosons. It seems that the $4$-body vanishing property does guarantee good variational energy, but absence of this property does not mean that the energy or overlap will necessarily be bad.

We will now make some remarks on future directions for this line of research.
All the work done here can be repeated for systems with an odd number of electrons (necessarily with different numbers of electrons in each layer). No results on multiplet counting have so far been published for such systems. For fermions, overlaps and spectra for the lowest lying neutral states will likely be similar to those presented in Ref.~\onlinecite{Sreejith11}. 

All constructions presented here can be generalized straigthforwardly to the Bonderson-Slingerland hierarchy states\cite{Bonderson08,Bonderson09}, which potentially describe all observed filling fractions for fermions in the second Landau level. The ground state wave functions of the Bonderson-Slingerland states contain the bosonic Pfaffian wave function as a factor; hence quasiparticles and excitons on this Pfaffian factor provide a natural set of trial wave functions for excitations of the Bonderson-Slingerland states. Other states, to which our construction generalizes naturally, are the non-Abelian condensate state of Ref. \cite{PhysRevLett.104.056803}. These can, in fact, be interpreted as  double layer states with a Jain CF state in each layer, even though their construction is via CFT.  The BCF states proposed in Ref.~\onlinecite{Sreejith11} are related to the states in Ref. \cite{PhysRevLett.104.056803} by a change of LLL-projection, in the same way that their pfaffian trial wave functions relate to the trial states presented here. 

We therefore expect, in line with the predictions of Ref. \cite{PhysRevLett.104.056803}, that they will have quasiparticle multiplet countings at large $L$ described by parafermionic CFTs.

Similarly, by working with $k$ composite fermion layers instead of just $2$ (still with symmetrisation over all layers), the construction presented here can be used to produce candidate quasiparticle and exciton wave functions for the entire series of Read-Rezayi states.\cite{Read99} When $k>2$, there is actually still an open question with regard to \emph{quasiholes} in this case; 
while it is very plausible\cite{PhysRevLett.101.066803}, it has not been strictly proven that the quasihole wave functions produced using the $k$-layer picture of these states and those that come from the description of the states in terms of vanishing properties are equivalent. That is, the layered wave functions satisfy the required vanishing properties, but is has not been proven that the angular momentum eigenstates which occur in their expansion span the full space of states with these vanishing properties. 

One may also define a version of the bilayer composite fermion states of~Ref.~\onlinecite{Sreejith11_PRL_107} which utilizes a LLL-projection more similar to the one used in the current work. The bosonic versions of these wave functions would simply be the symmetrized product of two composite fermion wave functions, where each of the composite fermion layers is at an effective filling $\nu^{*}=n$, with $n>1$ integer (the case $n=1$ with single flux attachment was studied in this paper). It would be of particular interest to find the counting of quasiparticle multiplets for these states. 

Another important direction for future work is to test our trial states against the spectra of more physically realistic Hamiltonians, often necessarily in larger Hilbert spaces. Here, one may think for instance of including the effects of the electron spin, Landau level mixing and subbands in finite thickness quantum wells.




\clearpage

\acknowledgments
\noindent JKS and IDR were supported by Science Foundation Ireland Principal Investigator award 08/IN.1/I1961.
MH was supported by the Alexander-von-Humboldt foundation, the Royal Swedish Academy of Science, and NSF DMR grant 0952428.

\bibliography{corr}

\end{document}